\documentclass[journal,onecolumn]{IEEEtrani}

\usepackage{caption}
\ifCLASSINFOpdf
\else
\fi
\usepackage{rotating}
\usepackage{url,hyperref,lineno,microtype,subcaption}
\usepackage[singlespacing]{setspace}
\usepackage{amsmath, amssymb}
\usepackage{booktabs}
\usepackage{pdflscape}
\usepackage{multirow}
\usepackage{algorithm}
\usepackage{algpseudocode}
\usepackage{comment}
\algrenewcommand\algorithmicrequire{\textbf{Input:}}
\algrenewcommand\algorithmicensure{\textbf{Output:}}

\usepackage[draft]{todonotes}
\usepackage{soul}

\usepackage{xspace}

\definecolor {snow}                {rgb}{1.00,0.98,0.98}
\definecolor {ghostwhite}          {rgb}{0.97,0.97,1.00}
\definecolor {whitesmoke}          {rgb}{0.96,0.96,0.96}
\definecolor {gainsboro}           {rgb}{0.86,0.86,0.86}
\definecolor {floralwhite}         {rgb}{1.00,0.98,0.94}
\definecolor {oldlace}             {rgb}{0.99,0.96,0.90}
\definecolor {linen}               {rgb}{0.98,0.94,0.90}
\definecolor {antiquewhite}        {rgb}{0.98,0.92,0.84}
\definecolor {papayawhip}          {rgb}{1.00,0.94,0.84}
\definecolor {blanchedalmond}      {rgb}{1.00,0.92,0.80}
\definecolor {bisque}              {rgb}{1.00,0.89,0.77}
\definecolor {peachpuff}           {rgb}{1.00,0.85,0.73}
\definecolor {navajowhite}         {rgb}{1.00,0.87,0.68}
\definecolor {moccasin}            {rgb}{1.00,0.89,0.71}
\definecolor {cornsilk}            {rgb}{1.00,0.97,0.86}
\definecolor {ivory}               {rgb}{1.00,1.00,0.94}
\definecolor {lemonchiffon}        {rgb}{1.00,0.98,0.80}
\definecolor {seashell}            {rgb}{1.00,0.96,0.93}
\definecolor {honeydew}            {rgb}{0.94,1.00,0.94}
\definecolor {mintcream}           {rgb}{0.96,1.00,0.98}
\definecolor {azure}               {rgb}{0.94,1.00,1.00}
\definecolor {aliceblue}           {rgb}{0.94,0.97,1.00}
\definecolor {lavender}            {rgb}{0.90,0.90,0.98}
\definecolor {lavenderblush}       {rgb}{1.00,0.94,0.96}
\definecolor {mistyrose}           {rgb}{1.00,0.89,0.88}
\definecolor {white}               {rgb}{1.00,1.00,1.00}
\definecolor {black}               {rgb}{0.00,0.00,0.00}
\definecolor {darkslategray}       {rgb}{0.18,0.31,0.31}
\definecolor {dimgray}             {rgb}{0.41,0.41,0.41}
\definecolor {slategray}           {rgb}{0.44,0.50,0.56}
\definecolor {lightslategray}      {rgb}{0.47,0.53,0.60}
\definecolor {gray}                {rgb}{0.75,0.75,0.75}
\definecolor {lightgrey}           {rgb}{0.83,0.83,0.83}
\definecolor {midnightblue}        {rgb}{0.10,0.10,0.44}
\definecolor {navy}                {rgb}{0.00,0.00,0.50}
\definecolor {cornflowerblue}      {rgb}{0.39,0.58,0.93}
\definecolor {darkslateblue}       {rgb}{0.28,0.24,0.55}
\definecolor {slateblue}           {rgb}{0.42,0.35,0.80}
\definecolor {mediumslateblue}     {rgb}{0.48,0.41,0.93}
\definecolor {lightslateblue}      {rgb}{0.52,0.44,1.00}
\definecolor {mediumblue}          {rgb}{0.00,0.00,0.80}
\definecolor {royalblue}           {rgb}{0.25,0.41,0.88}
\definecolor {blue}                {rgb}{0.00,0.00,1.00}
\definecolor {dodgerblue}          {rgb}{0.12,0.56,1.00}
\definecolor {deepskyblue}         {rgb}{0.00,0.75,1.00}
\definecolor {skyblue}             {rgb}{0.53,0.81,0.92}
\definecolor {lightskyblue}        {rgb}{0.53,0.81,0.98}
\definecolor {steelblue}           {rgb}{0.27,0.51,0.71}
\definecolor {lightsteelblue}      {rgb}{0.69,0.77,0.87}
\definecolor {lightblue}           {rgb}{0.68,0.85,0.90}
\definecolor {powderblue}          {rgb}{0.69,0.88,0.90}
\definecolor {paleturquoise}       {rgb}{0.69,0.93,0.93}
\definecolor {darkturquoise}       {rgb}{0.00,0.81,0.82}
\definecolor {mediumturquoise}     {rgb}{0.28,0.82,0.80}
\definecolor {turquoise}           {rgb}{0.25,0.88,0.82}
\definecolor {cyan}                {rgb}{0.00,1.00,1.00}
\definecolor {lightcyan}           {rgb}{0.88,1.00,1.00}
\definecolor {cadetblue}           {rgb}{0.37,0.62,0.63}
\definecolor {mediumaquamarine}    {rgb}{0.40,0.80,0.67}
\definecolor {aquamarine}          {rgb}{0.50,1.00,0.83}
\definecolor {darkgreen}           {rgb}{0.00,0.39,0.00}
\definecolor {darkolivegreen}      {rgb}{0.33,0.42,0.18}
\definecolor {darkseagreen}        {rgb}{0.56,0.74,0.56}
\definecolor {seagreen}            {rgb}{0.18,0.55,0.34}
\definecolor {mediumseagreen}      {rgb}{0.24,0.70,0.44}
\definecolor {lightseagreen}       {rgb}{0.13,0.70,0.67}
\definecolor {palegreen}           {rgb}{0.60,0.98,0.60}
\definecolor {springgreen}         {rgb}{0.00,1.00,0.50}
\definecolor {lawngreen}           {rgb}{0.49,0.99,0.00}
\definecolor {green}               {rgb}{0.00,1.00,0.00}
\definecolor {chartreuse}          {rgb}{0.50,1.00,0.00}
\definecolor {mediumspringgreen}   {rgb}{0.00,0.98,0.60}
\definecolor {greenyellow}         {rgb}{0.68,1.00,0.18}
\definecolor {limegreen}           {rgb}{0.20,0.80,0.20}
\definecolor {yellowgreen}         {rgb}{0.60,0.80,0.20}
\definecolor {forestgreen}         {rgb}{0.13,0.55,0.13}
\definecolor {olivedrab}           {rgb}{0.42,0.56,0.14}
\definecolor {darkkhaki}           {rgb}{0.74,0.72,0.42}
\definecolor {khaki}               {rgb}{0.94,0.90,0.55}
\definecolor {palegoldenrod}       {rgb}{0.93,0.91,0.67}
\definecolor {lightgoldenrodyellow} {rgb}{0.98,0.98,0.82}
\definecolor {lightyellow}         {rgb}{1.00,1.00,0.88}
\definecolor {yellow}              {rgb}{1.00,1.00,0.00}
\definecolor {gold}                {rgb}{1.00,0.84,0.00}
\definecolor {lightgoldenrod}      {rgb}{0.93,0.87,0.51}
\definecolor {goldenrod}           {rgb}{0.85,0.65,0.13}
\definecolor {darkgoldenrod}       {rgb}{0.72,0.53,0.04}
\definecolor {rosybrown}           {rgb}{0.74,0.56,0.56}
\definecolor {indianred}           {rgb}{0.80,0.36,0.36}
\definecolor {saddlebrown}         {rgb}{0.55,0.27,0.07}
\definecolor {sienna}              {rgb}{0.63,0.32,0.18}
\definecolor {peru}                {rgb}{0.80,0.52,0.25}
\definecolor {burlywood}           {rgb}{0.87,0.72,0.53}
\definecolor {beige}               {rgb}{0.96,0.96,0.86}
\definecolor {wheat}               {rgb}{0.96,0.87,0.70}
\definecolor {sandybrown}          {rgb}{0.96,0.64,0.38}
\definecolor {tan}                 {rgb}{0.82,0.71,0.55}
\definecolor {chocolate}           {rgb}{0.82,0.41,0.12}
\definecolor {firebrick}           {rgb}{0.70,0.13,0.13}
\definecolor {brown}               {rgb}{0.65,0.16,0.16}
\definecolor {darksalmon}          {rgb}{0.91,0.59,0.48}
\definecolor {salmon}              {rgb}{0.98,0.50,0.45}
\definecolor {lightsalmon}         {rgb}{1.00,0.63,0.48}
\definecolor {orange}              {rgb}{1.00,0.65,0.00}
\definecolor {darkorange}          {rgb}{1.00,0.55,0.00}
\definecolor {coral}               {rgb}{1.00,0.50,0.31}
\definecolor {lightcoral}          {rgb}{0.94,0.50,0.50}
\definecolor {tomato}              {rgb}{1.00,0.39,0.28}
\definecolor {orangered}           {rgb}{1.00,0.27,0.00}
\definecolor {red}                 {rgb}{1.00,0.00,0.00}
\definecolor {hotpink}             {rgb}{1.00,0.41,0.71}
\definecolor {deeppink}            {rgb}{1.00,0.08,0.58}
\definecolor {pink}                {rgb}{1.00,0.75,0.80}
\definecolor {lightpink}           {rgb}{1.00,0.71,0.76}
\definecolor {palevioletred}       {rgb}{0.86,0.44,0.58}
\definecolor {maroon}              {rgb}{0.69,0.19,0.38}
\definecolor {mediumvioletred}     {rgb}{0.78,0.08,0.52}
\definecolor {violetred}           {rgb}{0.82,0.13,0.56}
\definecolor {magenta}             {rgb}{1.00,0.00,1.00}
\definecolor {violet}              {rgb}{0.93,0.51,0.93}
\definecolor {plum}                {rgb}{0.87,0.63,0.87}
\definecolor {orchid}              {rgb}{0.85,0.44,0.84}
\definecolor {mediumorchid}        {rgb}{0.73,0.33,0.83}
\definecolor {darkorchid}          {rgb}{0.60,0.20,0.80}
\definecolor {darkviolet}          {rgb}{0.58,0.00,0.83}
\definecolor {blueviolet}          {rgb}{0.54,0.17,0.89}
\definecolor {purple}              {rgb}{0.63,0.13,0.94}
\definecolor {mediumpurple}        {rgb}{0.58,0.44,0.86}
\definecolor {thistle}             {rgb}{0.85,0.75,0.85}
\definecolor {snow2}               {rgb}{0.93,0.91,0.91}
\definecolor {snow3}               {rgb}{0.80,0.79,0.79}
\definecolor {snow4}               {rgb}{0.55,0.54,0.54}
\definecolor {seashell2}           {rgb}{0.93,0.90,0.87}
\definecolor {seashell3}           {rgb}{0.80,0.77,0.75}
\definecolor {seashell4}           {rgb}{0.55,0.53,0.51}
\definecolor {antiquewhite1}       {rgb}{1.00,0.94,0.86}
\definecolor {antiquewhite2}       {rgb}{0.93,0.87,0.80}
\definecolor {antiquewhite3}       {rgb}{0.80,0.75,0.69}
\definecolor {antiquewhite4}       {rgb}{0.55,0.51,0.47}
\definecolor {bisque2}             {rgb}{0.93,0.84,0.72}
\definecolor {bisque3}             {rgb}{0.80,0.72,0.62}
\definecolor {bisque4}             {rgb}{0.55,0.49,0.42}
\definecolor {peachpuff2}          {rgb}{0.93,0.80,0.68}
\definecolor {peachpuff3}          {rgb}{0.80,0.69,0.58}
\definecolor {peachpuff4}          {rgb}{0.55,0.47,0.40}
\definecolor {navajowhite2}        {rgb}{0.93,0.81,0.63}
\definecolor {navajowhite3}        {rgb}{0.80,0.70,0.55}
\definecolor {navajowhite4}        {rgb}{0.55,0.47,0.37}
\definecolor {lemonchiffon2}       {rgb}{0.93,0.91,0.75}
\definecolor {lemonchiffon3}       {rgb}{0.80,0.79,0.65}
\definecolor {lemonchiffon4}       {rgb}{0.55,0.54,0.44}
\definecolor {cornsilk2}           {rgb}{0.93,0.91,0.80}
\definecolor {cornsilk3}           {rgb}{0.80,0.78,0.69}
\definecolor {cornsilk4}           {rgb}{0.55,0.53,0.47}
\definecolor {ivory2}              {rgb}{0.93,0.93,0.88}
\definecolor {ivory3}              {rgb}{0.80,0.80,0.76}
\definecolor {ivory4}              {rgb}{0.55,0.55,0.51}
\definecolor {honeydew2}           {rgb}{0.88,0.93,0.88}
\definecolor {honeydew3}           {rgb}{0.76,0.80,0.76}
\definecolor {honeydew4}           {rgb}{0.51,0.55,0.51}
\definecolor {lavenderblush2}      {rgb}{0.93,0.88,0.90}
\definecolor {lavenderblush3}      {rgb}{0.80,0.76,0.77}
\definecolor {lavenderblush4}      {rgb}{0.55,0.51,0.53}
\definecolor {mistyrose2}          {rgb}{0.93,0.84,0.82}
\definecolor {mistyrose3}          {rgb}{0.80,0.72,0.71}
\definecolor {mistyrose4}          {rgb}{0.55,0.49,0.48}
\definecolor {azure2}              {rgb}{0.88,0.93,0.93}
\definecolor {azure3}              {rgb}{0.76,0.80,0.80}
\definecolor {azure4}              {rgb}{0.51,0.55,0.55}
\definecolor {slateblue1}          {rgb}{0.51,0.44,1.00}
\definecolor {slateblue2}          {rgb}{0.48,0.40,0.93}
\definecolor {slateblue3}          {rgb}{0.41,0.35,0.80}
\definecolor {slateblue4}          {rgb}{0.28,0.24,0.55}
\definecolor {royalblue1}          {rgb}{0.28,0.46,1.00}
\definecolor {royalblue2}          {rgb}{0.26,0.43,0.93}
\definecolor {royalblue3}          {rgb}{0.23,0.37,0.80}
\definecolor {royalblue4}          {rgb}{0.15,0.25,0.55}
\definecolor {blue2}               {rgb}{0.00,0.00,0.93}
\definecolor {blue4}               {rgb}{0.00,0.00,0.55}
\definecolor {dodgerblue2}         {rgb}{0.11,0.53,0.93}
\definecolor {dodgerblue3}         {rgb}{0.09,0.45,0.80}
\definecolor {dodgerblue4}         {rgb}{0.06,0.31,0.55}
\definecolor {steelblue1}          {rgb}{0.39,0.72,1.00}
\definecolor {steelblue2}          {rgb}{0.36,0.67,0.93}
\definecolor {steelblue3}          {rgb}{0.31,0.58,0.80}
\definecolor {steelblue4}          {rgb}{0.21,0.39,0.55}
\definecolor {deepskyblue2}        {rgb}{0.00,0.70,0.93}
\definecolor {deepskyblue3}        {rgb}{0.00,0.60,0.80}
\definecolor {deepskyblue4}        {rgb}{0.00,0.41,0.55}
\definecolor {skyblue1}            {rgb}{0.53,0.81,1.00}
\definecolor {skyblue2}            {rgb}{0.49,0.75,0.93}
\definecolor {skyblue3}            {rgb}{0.42,0.65,0.80}
\definecolor {skyblue4}            {rgb}{0.29,0.44,0.55}
\definecolor {lightskyblue1}       {rgb}{0.69,0.89,1.00}
\definecolor {lightskyblue2}       {rgb}{0.64,0.83,0.93}
\definecolor {lightskyblue3}       {rgb}{0.55,0.71,0.80}
\definecolor {lightskyblue4}       {rgb}{0.38,0.48,0.55}
\definecolor {slategray1}          {rgb}{0.78,0.89,1.00}
\definecolor {slategray2}          {rgb}{0.73,0.83,0.93}
\definecolor {slategray3}          {rgb}{0.62,0.71,0.80}
\definecolor {slategray4}          {rgb}{0.42,0.48,0.55}
\definecolor {lightsteelblue1}     {rgb}{0.79,0.88,1.00}
\definecolor {lightsteelblue2}     {rgb}{0.74,0.82,0.93}
\definecolor {lightsteelblue3}     {rgb}{0.64,0.71,0.80}
\definecolor {lightsteelblue4}     {rgb}{0.43,0.48,0.55}
\definecolor {lightblue1}          {rgb}{0.75,0.94,1.00}
\definecolor {lightblue2}          {rgb}{0.70,0.87,0.93}
\definecolor {lightblue3}          {rgb}{0.60,0.75,0.80}
\definecolor {lightblue4}          {rgb}{0.41,0.51,0.55}
\definecolor {lightcyan2}          {rgb}{0.82,0.93,0.93}
\definecolor {lightcyan3}          {rgb}{0.71,0.80,0.80}
\definecolor {lightcyan4}          {rgb}{0.48,0.55,0.55}
\definecolor {paleturquoise1}      {rgb}{0.73,1.00,1.00}
\definecolor {paleturquoise2}      {rgb}{0.68,0.93,0.93}
\definecolor {paleturquoise3}      {rgb}{0.59,0.80,0.80}
\definecolor {paleturquoise4}      {rgb}{0.40,0.55,0.55}
\definecolor {cadetblue1}          {rgb}{0.60,0.96,1.00}
\definecolor {cadetblue2}          {rgb}{0.56,0.90,0.93}
\definecolor {cadetblue3}          {rgb}{0.48,0.77,0.80}
\definecolor {cadetblue4}          {rgb}{0.33,0.53,0.55}
\definecolor {turquoise1}          {rgb}{0.00,0.96,1.00}
\definecolor {turquoise2}          {rgb}{0.00,0.90,0.93}
\definecolor {turquoise3}          {rgb}{0.00,0.77,0.80}
\definecolor {turquoise4}          {rgb}{0.00,0.53,0.55}
\definecolor {cyan2}               {rgb}{0.00,0.93,0.93}
\definecolor {cyan3}               {rgb}{0.00,0.80,0.80}
\definecolor {cyan4}               {rgb}{0.00,0.55,0.55}
\definecolor {darkslategray1}      {rgb}{0.59,1.00,1.00}
\definecolor {darkslategray2}      {rgb}{0.55,0.93,0.93}
\definecolor {darkslategray3}      {rgb}{0.47,0.80,0.80}
\definecolor {darkslategray4}      {rgb}{0.32,0.55,0.55}
\definecolor {aquamarine2}         {rgb}{0.46,0.93,0.78}
\definecolor {aquamarine4}         {rgb}{0.27,0.55,0.45}
\definecolor {darkseagreen1}       {rgb}{0.76,1.00,0.76}
\definecolor {darkseagreen2}       {rgb}{0.71,0.93,0.71}
\definecolor {darkseagreen3}       {rgb}{0.61,0.80,0.61}
\definecolor {darkseagreen4}       {rgb}{0.41,0.55,0.41}
\definecolor {seagreen1}           {rgb}{0.33,1.00,0.62}
\definecolor {seagreen2}           {rgb}{0.31,0.93,0.58}
\definecolor {seagreen3}           {rgb}{0.26,0.80,0.50}
\definecolor {palegreen1}          {rgb}{0.60,1.00,0.60}
\definecolor {palegreen2}          {rgb}{0.56,0.93,0.56}
\definecolor {palegreen3}          {rgb}{0.49,0.80,0.49}
\definecolor {palegreen4}          {rgb}{0.33,0.55,0.33}
\definecolor {springgreen2}        {rgb}{0.00,0.93,0.46}
\definecolor {springgreen3}        {rgb}{0.00,0.80,0.40}
\definecolor {springgreen4}        {rgb}{0.00,0.55,0.27}
\definecolor {green2}              {rgb}{0.00,0.93,0.00}
\definecolor {green3}              {rgb}{0.00,0.80,0.00}
\definecolor {green4}              {rgb}{0.00,0.55,0.00}
\definecolor {chartreuse2}         {rgb}{0.46,0.93,0.00}
\definecolor {chartreuse3}         {rgb}{0.40,0.80,0.00}
\definecolor {chartreuse4}         {rgb}{0.27,0.55,0.00}
\definecolor {olivedrab1}          {rgb}{0.75,1.00,0.24}
\definecolor {olivedrab2}          {rgb}{0.70,0.93,0.23}
\definecolor {olivedrab4}          {rgb}{0.41,0.55,0.13}
\definecolor {darkolivegreen1}     {rgb}{0.79,1.00,0.44}
\definecolor {darkolivegreen2}     {rgb}{0.74,0.93,0.41}
\definecolor {darkolivegreen3}     {rgb}{0.64,0.80,0.35}
\definecolor {darkolivegreen4}     {rgb}{0.43,0.55,0.24}
\definecolor {khaki1}              {rgb}{1.00,0.96,0.56}
\definecolor {khaki2}              {rgb}{0.93,0.90,0.52}
\definecolor {khaki3}              {rgb}{0.80,0.78,0.45}
\definecolor {khaki4}              {rgb}{0.55,0.53,0.31}
\definecolor {lightgoldenrod1}     {rgb}{1.00,0.93,0.55}
\definecolor {lightgoldenrod2}     {rgb}{0.93,0.86,0.51}
\definecolor {lightgoldenrod3}     {rgb}{0.80,0.75,0.44}
\definecolor {lightgoldenrod4}     {rgb}{0.55,0.51,0.30}
\definecolor {lightyellow2}        {rgb}{0.93,0.93,0.82}
\definecolor {lightyellow3}        {rgb}{0.80,0.80,0.71}
\definecolor {lightyellow4}        {rgb}{0.55,0.55,0.48}
\definecolor {yellow2}             {rgb}{0.93,0.93,0.00}
\definecolor {yellow3}             {rgb}{0.80,0.80,0.00}
\definecolor {yellow4}             {rgb}{0.55,0.55,0.00}
\definecolor {gold2}               {rgb}{0.93,0.79,0.00}
\definecolor {gold3}               {rgb}{0.80,0.68,0.00}
\definecolor {gold4}               {rgb}{0.55,0.46,0.00}
\definecolor {goldenrod1}          {rgb}{1.00,0.76,0.15}
\definecolor {goldenrod2}          {rgb}{0.93,0.71,0.13}
\definecolor {goldenrod3}          {rgb}{0.80,0.61,0.11}
\definecolor {goldenrod4}          {rgb}{0.55,0.41,0.08}
\definecolor {darkgoldenrod1}      {rgb}{1.00,0.73,0.06}
\definecolor {darkgoldenrod2}      {rgb}{0.93,0.68,0.05}
\definecolor {darkgoldenrod3}      {rgb}{0.80,0.58,0.05}
\definecolor {darkgoldenrod4}      {rgb}{0.55,0.40,0.03}
\definecolor {rosybrown1}          {rgb}{1.00,0.76,0.76}
\definecolor {rosybrown2}          {rgb}{0.93,0.71,0.71}
\definecolor {rosybrown3}          {rgb}{0.80,0.61,0.61}
\definecolor {rosybrown4}          {rgb}{0.55,0.41,0.41}
\definecolor {indianred1}          {rgb}{1.00,0.42,0.42}
\definecolor {indianred2}          {rgb}{0.93,0.39,0.39}
\definecolor {indianred3}          {rgb}{0.80,0.33,0.33}
\definecolor {indianred4}          {rgb}{0.55,0.23,0.23}
\definecolor {sienna1}             {rgb}{1.00,0.51,0.28}
\definecolor {sienna2}             {rgb}{0.93,0.47,0.26}
\definecolor {sienna3}             {rgb}{0.80,0.41,0.22}
\definecolor {sienna4}             {rgb}{0.55,0.28,0.15}
\definecolor {burlywood1}          {rgb}{1.00,0.83,0.61}
\definecolor {burlywood2}          {rgb}{0.93,0.77,0.57}
\definecolor {burlywood3}          {rgb}{0.80,0.67,0.49}
\definecolor {burlywood4}          {rgb}{0.55,0.45,0.33}
\definecolor {wheat1}              {rgb}{1.00,0.91,0.73}
\definecolor {wheat2}              {rgb}{0.93,0.85,0.68}
\definecolor {wheat3}              {rgb}{0.80,0.73,0.59}
\definecolor {wheat4}              {rgb}{0.55,0.49,0.40}
\definecolor {tan1}                {rgb}{1.00,0.65,0.31}
\definecolor {tan2}                {rgb}{0.93,0.60,0.29}
\definecolor {tan4}                {rgb}{0.55,0.35,0.17}
\definecolor {chocolate1}          {rgb}{1.00,0.50,0.14}
\definecolor {chocolate2}          {rgb}{0.93,0.46,0.13}
\definecolor {chocolate3}          {rgb}{0.80,0.40,0.11}
\definecolor {firebrick1}          {rgb}{1.00,0.19,0.19}
\definecolor {firebrick2}          {rgb}{0.93,0.17,0.17}
\definecolor {firebrick3}          {rgb}{0.80,0.15,0.15}
\definecolor {firebrick4}          {rgb}{0.55,0.10,0.10}
\definecolor {brown1}              {rgb}{1.00,0.25,0.25}
\definecolor {brown2}              {rgb}{0.93,0.23,0.23}
\definecolor {brown3}              {rgb}{0.80,0.20,0.20}
\definecolor {brown4}              {rgb}{0.55,0.14,0.14}
\definecolor {salmon1}             {rgb}{1.00,0.55,0.41}
\definecolor {salmon2}             {rgb}{0.93,0.51,0.38}
\definecolor {salmon3}             {rgb}{0.80,0.44,0.33}
\definecolor {salmon4}             {rgb}{0.55,0.30,0.22}
\definecolor {lightsalmon2}        {rgb}{0.93,0.58,0.45}
\definecolor {lightsalmon3}        {rgb}{0.80,0.51,0.38}
\definecolor {lightsalmon4}        {rgb}{0.55,0.34,0.26}
\definecolor {orange2}             {rgb}{0.93,0.60,0.00}
\definecolor {orange3}             {rgb}{0.80,0.52,0.00}
\definecolor {orange4}             {rgb}{0.55,0.35,0.00}
\definecolor {darkorange1}         {rgb}{1.00,0.50,0.00}
\definecolor {darkorange2}         {rgb}{0.93,0.46,0.00}
\definecolor {darkorange3}         {rgb}{0.80,0.40,0.00}
\definecolor {darkorange4}         {rgb}{0.55,0.27,0.00}
\definecolor {coral1}              {rgb}{1.00,0.45,0.34}
\definecolor {coral2}              {rgb}{0.93,0.42,0.31}
\definecolor {coral3}              {rgb}{0.80,0.36,0.27}
\definecolor {coral4}              {rgb}{0.55,0.24,0.18}
\definecolor {tomato2}             {rgb}{0.93,0.36,0.26}
\definecolor {tomato3}             {rgb}{0.80,0.31,0.22}
\definecolor {tomato4}             {rgb}{0.55,0.21,0.15}
\definecolor {orangered2}          {rgb}{0.93,0.25,0.00}
\definecolor {orangered3}          {rgb}{0.80,0.22,0.00}
\definecolor {orangered4}          {rgb}{0.55,0.15,0.00}
\definecolor {red2}                {rgb}{0.93,0.00,0.00}
\definecolor {red3}                {rgb}{0.80,0.00,0.00}
\definecolor {red4}                {rgb}{0.55,0.00,0.00}
\definecolor {deeppink2}           {rgb}{0.93,0.07,0.54}
\definecolor {deeppink3}           {rgb}{0.80,0.06,0.46}
\definecolor {deeppink4}           {rgb}{0.55,0.04,0.31}
\definecolor {hotpink1}            {rgb}{1.00,0.43,0.71}
\definecolor {hotpink2}            {rgb}{0.93,0.42,0.65}
\definecolor {hotpink3}            {rgb}{0.80,0.38,0.56}
\definecolor {hotpink4}            {rgb}{0.55,0.23,0.38}
\definecolor {pink1}               {rgb}{1.00,0.71,0.77}
\definecolor {pink2}               {rgb}{0.93,0.66,0.72}
\definecolor {pink3}               {rgb}{0.80,0.57,0.62}
\definecolor {pink4}               {rgb}{0.55,0.39,0.42}
\definecolor {lightpink1}          {rgb}{1.00,0.68,0.73}
\definecolor {lightpink2}          {rgb}{0.93,0.64,0.68}
\definecolor {lightpink3}          {rgb}{0.80,0.55,0.58}
\definecolor {lightpink4}          {rgb}{0.55,0.37,0.40}
\definecolor {palevioletred1}      {rgb}{1.00,0.51,0.67}
\definecolor {palevioletred2}      {rgb}{0.93,0.47,0.62}
\definecolor {palevioletred3}      {rgb}{0.80,0.41,0.54}
\definecolor {palevioletred4}      {rgb}{0.55,0.28,0.36}
\definecolor {maroon1}             {rgb}{1.00,0.20,0.70}
\definecolor {maroon2}             {rgb}{0.93,0.19,0.65}
\definecolor {maroon3}             {rgb}{0.80,0.16,0.56}
\definecolor {maroon4}             {rgb}{0.55,0.11,0.38}
\definecolor {violetred1}          {rgb}{1.00,0.24,0.59}
\definecolor {violetred2}          {rgb}{0.93,0.23,0.55}
\definecolor {violetred3}          {rgb}{0.80,0.20,0.47}
\definecolor {violetred4}          {rgb}{0.55,0.13,0.32}
\definecolor {magenta2}            {rgb}{0.93,0.00,0.93}
\definecolor {magenta3}            {rgb}{0.80,0.00,0.80}
\definecolor {magenta4}            {rgb}{0.55,0.00,0.55}
\definecolor {orchid1}             {rgb}{1.00,0.51,0.98}
\definecolor {orchid2}             {rgb}{0.93,0.48,0.91}
\definecolor {orchid3}             {rgb}{0.80,0.41,0.79}
\definecolor {orchid4}             {rgb}{0.55,0.28,0.54}
\definecolor {plum1}               {rgb}{1.00,0.73,1.00}
\definecolor {plum2}               {rgb}{0.93,0.68,0.93}
\definecolor {plum3}               {rgb}{0.80,0.59,0.80}
\definecolor {plum4}               {rgb}{0.55,0.40,0.55}
\definecolor {mediumorchid1}       {rgb}{0.88,0.40,1.00}
\definecolor {mediumorchid2}       {rgb}{0.82,0.37,0.93}
\definecolor {mediumorchid3}       {rgb}{0.71,0.32,0.80}
\definecolor {mediumorchid4}       {rgb}{0.48,0.22,0.55}
\definecolor {darkorchid1}         {rgb}{0.75,0.24,1.00}
\definecolor {darkorchid2}         {rgb}{0.70,0.23,0.93}
\definecolor {darkorchid3}         {rgb}{0.60,0.20,0.80}
\definecolor {darkorchid4}         {rgb}{0.41,0.13,0.55}
\definecolor {purple1}             {rgb}{0.61,0.19,1.00}
\definecolor {purple2}             {rgb}{0.57,0.17,0.93}
\definecolor {purple3}             {rgb}{0.49,0.15,0.80}
\definecolor {purple4}             {rgb}{0.33,0.10,0.55}
\definecolor {mediumpurple1}       {rgb}{0.67,0.51,1.00}
\definecolor {mediumpurple2}       {rgb}{0.62,0.47,0.93}
\definecolor {mediumpurple3}       {rgb}{0.54,0.41,0.80}
\definecolor {mediumpurple4}       {rgb}{0.36,0.28,0.55}
\definecolor {thistle1}            {rgb}{1.00,0.88,1.00}
\definecolor {thistle2}            {rgb}{0.93,0.82,0.93}
\definecolor {thistle3}            {rgb}{0.80,0.71,0.80}
\definecolor {thistle4}            {rgb}{0.55,0.48,0.55}
\definecolor {gray1}               {rgb}{0.01,0.01,0.01}
\definecolor {gray2}               {rgb}{0.02,0.02,0.02}
\definecolor {gray3}               {rgb}{0.03,0.03,0.03}
\definecolor {gray4}               {rgb}{0.04,0.04,0.04}
\definecolor {gray5}               {rgb}{0.05,0.05,0.05}
\definecolor {gray6}               {rgb}{0.06,0.06,0.06}
\definecolor {gray7}               {rgb}{0.07,0.07,0.07}
\definecolor {gray8}               {rgb}{0.08,0.08,0.08}
\definecolor {gray9}               {rgb}{0.09,0.09,0.09}
\definecolor {gray10}              {rgb}{0.10,0.10,0.10}
\definecolor {gray11}              {rgb}{0.11,0.11,0.11}
\definecolor {gray12}              {rgb}{0.12,0.12,0.12}
\definecolor {gray13}              {rgb}{0.13,0.13,0.13}
\definecolor {gray14}              {rgb}{0.14,0.14,0.14}
\definecolor {gray15}              {rgb}{0.15,0.15,0.15}
\definecolor {gray16}              {rgb}{0.16,0.16,0.16}
\definecolor {gray17}              {rgb}{0.17,0.17,0.17}
\definecolor {gray18}              {rgb}{0.18,0.18,0.18}
\definecolor {gray19}              {rgb}{0.19,0.19,0.19}
\definecolor {gray20}              {rgb}{0.20,0.20,0.20}
\definecolor {gray21}              {rgb}{0.21,0.21,0.21}
\definecolor {gray22}              {rgb}{0.22,0.22,0.22}
\definecolor {gray23}              {rgb}{0.23,0.23,0.23}
\definecolor {gray24}              {rgb}{0.24,0.24,0.24}
\definecolor {gray25}              {rgb}{0.25,0.25,0.25}
\definecolor {gray26}              {rgb}{0.26,0.26,0.26}
\definecolor {gray27}              {rgb}{0.27,0.27,0.27}
\definecolor {gray28}              {rgb}{0.28,0.28,0.28}
\definecolor {gray29}              {rgb}{0.29,0.29,0.29}
\definecolor {gray30}              {rgb}{0.30,0.30,0.30}
\definecolor {gray31}              {rgb}{0.31,0.31,0.31}
\definecolor {gray32}              {rgb}{0.32,0.32,0.32}
\definecolor {gray33}              {rgb}{0.33,0.33,0.33}
\definecolor {gray34}              {rgb}{0.34,0.34,0.34}
\definecolor {gray35}              {rgb}{0.35,0.35,0.35}
\definecolor {gray36}              {rgb}{0.36,0.36,0.36}
\definecolor {gray37}              {rgb}{0.37,0.37,0.37}
\definecolor {gray38}              {rgb}{0.38,0.38,0.38}
\definecolor {gray39}              {rgb}{0.39,0.39,0.39}
\definecolor {gray40}              {rgb}{0.40,0.40,0.40}
\definecolor {gray42}              {rgb}{0.42,0.42,0.42}
\definecolor {gray43}              {rgb}{0.43,0.43,0.43}
\definecolor {gray44}              {rgb}{0.44,0.44,0.44}
\definecolor {gray45}              {rgb}{0.45,0.45,0.45}
\definecolor {gray46}              {rgb}{0.46,0.46,0.46}
\definecolor {gray47}              {rgb}{0.47,0.47,0.47}
\definecolor {gray48}              {rgb}{0.48,0.48,0.48}
\definecolor {gray49}              {rgb}{0.49,0.49,0.49}
\definecolor {gray50}              {rgb}{0.50,0.50,0.50}
\definecolor {gray51}              {rgb}{0.51,0.51,0.51}
\definecolor {gray52}              {rgb}{0.52,0.52,0.52}
\definecolor {gray53}              {rgb}{0.53,0.53,0.53}
\definecolor {gray54}              {rgb}{0.54,0.54,0.54}
\definecolor {gray55}              {rgb}{0.55,0.55,0.55}
\definecolor {gray56}              {rgb}{0.56,0.56,0.56}
\definecolor {gray57}              {rgb}{0.57,0.57,0.57}
\definecolor {gray58}              {rgb}{0.58,0.58,0.58}
\definecolor {gray59}              {rgb}{0.59,0.59,0.59}
\definecolor {gray60}              {rgb}{0.60,0.60,0.60}
\definecolor {gray61}              {rgb}{0.61,0.61,0.61}
\definecolor {gray62}              {rgb}{0.62,0.62,0.62}
\definecolor {gray63}              {rgb}{0.63,0.63,0.63}
\definecolor {gray64}              {rgb}{0.64,0.64,0.64}
\definecolor {gray65}              {rgb}{0.65,0.65,0.65}
\definecolor {gray66}              {rgb}{0.66,0.66,0.66}
\definecolor {gray67}              {rgb}{0.67,0.67,0.67}
\definecolor {gray68}              {rgb}{0.68,0.68,0.68}
\definecolor {gray69}              {rgb}{0.69,0.69,0.69}
\definecolor {gray70}              {rgb}{0.70,0.70,0.70}
\definecolor {gray71}              {rgb}{0.71,0.71,0.71}
\definecolor {gray72}              {rgb}{0.72,0.72,0.72}
\definecolor {gray73}              {rgb}{0.73,0.73,0.73}
\definecolor {gray74}              {rgb}{0.74,0.74,0.74}
\definecolor {gray75}              {rgb}{0.75,0.75,0.75}
\definecolor {gray76}              {rgb}{0.76,0.76,0.76}
\definecolor {gray77}              {rgb}{0.77,0.77,0.77}
\definecolor {gray78}              {rgb}{0.78,0.78,0.78}
\definecolor {gray79}              {rgb}{0.79,0.79,0.79}
\definecolor {gray80}              {rgb}{0.80,0.80,0.80}
\definecolor {gray81}              {rgb}{0.81,0.81,0.81}
\definecolor {gray82}              {rgb}{0.82,0.82,0.82}
\definecolor {gray83}              {rgb}{0.83,0.83,0.83}
\definecolor {gray84}              {rgb}{0.84,0.84,0.84}
\definecolor {gray85}              {rgb}{0.85,0.85,0.85}
\definecolor {gray86}              {rgb}{0.86,0.86,0.86}
\definecolor {gray87}              {rgb}{0.87,0.87,0.87}
\definecolor {gray88}              {rgb}{0.88,0.88,0.88}
\definecolor {gray89}              {rgb}{0.89,0.89,0.89}
\definecolor {gray90}              {rgb}{0.90,0.90,0.90}
\definecolor {gray91}              {rgb}{0.91,0.91,0.91}
\definecolor {gray92}              {rgb}{0.92,0.92,0.92}
\definecolor {gray93}              {rgb}{0.93,0.93,0.93}
\definecolor {gray94}              {rgb}{0.94,0.94,0.94}
\definecolor {gray95}              {rgb}{0.95,0.95,0.95}
\definecolor {gray97}              {rgb}{0.97,0.97,0.97}
\definecolor {gray98}              {rgb}{0.98,0.98,0.98}
\definecolor {gray99}              {rgb}{0.99,0.99,0.99}
\definecolor {darkgrey}            {rgb}{0.66,0.66,0.66}

\newcommand{\offset}{\ensuremath{{\sf o}}}
\renewcommand{\offset}{\ensuremath{\theta_{0}}}

\newcommand{\xs}{\ensuremath{\underline{\mathbf{x}}}\xspace}

\newcommand{\zs}{\ensuremath{\underline{\mathbf{z}}}\xspace}

\newcommand{\as}{\ensuremath{\underline{\mathbf{a}}}\xspace}

\newcommand{\ts}{}
\renewcommand{\ts}{\ensuremath{\underline{\boldsymbol{\theta}}}\xspace}
\newcommand{\Fx}{\ensuremath{F(\xs)}\xspace}

\newcommand{\Px}{\ensuremath{P_F(\xs|\ts)}\xspace}

\newcommand{\Pxa}{\ensuremath{P_F(\xs,\as|\ts)}\xspace}

\newcommand{\ignoreinshort}[1]{}

\newcommand{\Pxaz}{\ensuremath{P_F(\underbrace{\xs,\as}_{\zs}|\ts)}\xspace}
\renewcommand{\Pxaz}{\ensuremath{P_F(\overbrace{\xs,\as}^{\zs}|\ts)}\xspace}

\newcommand{\TODO}[1]{{}}

\newcommand{\ignore}[1]{}
\newcommand{\RSCHANGE}[1]{\textcolor{blue}{#1}}

\newcommand{\RSTODO}[1]{{\bf \textcolor{darkgreen}{{\fbox{RS TODO:} #1}}}}
\renewcommand{\RSTODO}[1]{}
\def\makenewenumerate#1#2{%
\newcounter{cnt#1}
\newenvironment{#1}%
{\begin{list}{\makebox[0pt][r]{#2}}%
{\setlength{\itemsep}{0pt}% 
 \setlength{\parsep}{.2em}%
 \setlength{\leftmargin}{1.5em}%
 \setlength{\labelwidth}{.4em}%
 \usecounter{cnt#1}}}
{\end{list}}}

\makenewenumerate{myenumerate}{\arabic{cntmyenumerate}.}

\makenewenumerate{renumerate}{\rm(\roman{cntrenumerate})}
\makenewenumerate{renumerateprime}{\rm(\roman{cntrenumerateprime}$'$)}
\makenewenumerate{renumeratesecond}{\rm(\roman{cntrenumeratesecond}$''$)}

\makenewenumerate{aenumerate}{({\it\alph{cntaenumerate}})}
\makenewenumerate{aenumerateprime}{({\it\alph{cntaenumerateprime}$'$})}
\makenewenumerate{aenumeratesecond}{({\it\alph{cntaenumeratesecond}$''$})}

\def\newplaintheorem#1#2{%
\newtheorem{#1plain}{#2}% %% RS: mon mi piace l'indice di sezione
\newenvironment{#1}{\begin{#1plain}\rm }{\end{#1plain}}}

\newplaintheorem{notation}{Notation}
\newplaintheorem{cexample}{Counter-Example}

\newcommand{\tuple}[1]{\ensuremath{\langle{#1}\rangle}\xspace}
\newcommand{\set}[1]{\ensuremath{\{{#1}\}}\xspace}

\newcommand{\defas}{\ensuremath{\stackrel{\text{\tiny def}}{=}}\xspace}

\renewcommand{\RSCHANGE}[1]{\textcolor{black}{{#1}}}
\newcommand\mysout{\bgroup \markoverwith{{-}}\ULon}
\newcommand\nosout{\bgroup \markoverwith{{ }}\ULon}
\definecolor{mygray}{rgb}{0.90,0.90,0.90}
\definecolor{mywhite}{rgb}{1.00,1.00,1.00}

\newcommand{\optimathsat}{\textsc{OptiMathSAT}\xspace}

\renewcommand{\RSTODO}[1]{\noindent{\textcolor{blue}{{\fbox{RS TODO:} #1}}}}

\renewcommand{\TODO}[1]{\noindent{\textcolor{darkviolet}{{\fbox{TODO:} #1}}}}

\newcommand{\ti}[1]{\ensuremath{\sf{t}^{(#1)}}\xspace}
\newcommand{\tn}[1]{\ti{n}}

\renewcommand{\TODO}[1]{\todo[inline,color=green!40]{{\small{#1}}}}

\renewcommand{\RSTODO}[1]{\todo[inline,color=green!40]{{\small{RS TODO: #1}}}}

\newcommand{\GMCHANGE}[1]{\textcolor{black}{#1}}

\newcommand{\myf}[1]{f_{#1}}

\renewcommand{\myf}[1]
{\ifthenelse%
 {\equal{#1}{11}}{x_1^2x_2}%
 {\ifthenelse%
  {\equal{#1}{12}}{x_1^3x_2}
  {\ifthenelse%
   {\equal{#1}{21}}{x_1x_2^2}
   {\ifthenelse%
    {\equal{#1}{22}}{x_1x_2^3}
    {\ifthenelse%
     {\equal{#1}{3}}{2x_1x_2}
     {3x_1x_2}
}}}}}

\newcommand{\OptFN}[2]{{\ifx&#2&\ensuremath{#1}\else\ensuremath{#1(#2)}\fi}}

\newcommand{\JDst}[1]{{\color{darkviolet} \st{#1}}}
\newcommand{\jdREVISION}[1]{{\color{black} #1}}

\renewcommand{\JDst}[1]{}
\usepackage[backend=biber,style=numeric,sorting=nyt,maxbibnames=3,minbibnames=1]{biblatex} % Add the biblatex package
\usepackage{placeins}

\begin{document}
%
% paper title
% Titles are generally capitalized except for words such as a, an, and, as,
% at, but, by, for, in, nor, of, on, or, the, to and up, which are usually
% not capitalized unless they are the first or last word of the title.
% Linebreaks \\ can be used within to get better formatting as desired.
% Do not put math or special symbols in the title.
\title{Experimenting with D-Wave Quantum Annealers 
on Prime Factorization problems}
%
%
% author names and IEEE memberships
% note positions of commas and nonbreaking spaces ( ~ ) LaTeX will not break
% a structure at a ~ so this keeps an author's name from being broken across
% two lines.
% use \thanks{} to gain access to the first footnote area
% a separate \thanks must be used for each paragraph as LaTeX2e's \thanks
% was not built to handle multiple paragraphs
%

\author{Jingwen~Ding,
        Giuseppe~Spallitta,
        ~Roberto~Sebastiani% <-this % stops a space
\thanks{}% <-this % stops a space
\thanks{}% <-this % stops a space
\thanks{Manuscript published on Frontiers.}}

% note the % following the last \IEEEmembership and also \thanks - 
% these prevent an unwanted space from occurring between the last author name
% and the end of the author line. i.e., if you had this:
% 
% \author{....lastname \thanks{...} \thanks{...} }
%                     ^------------^------------^----Do not want these spaces!
%
% a space would be appended to the last name and could cause every name on that
% line to be shifted left slightly. This is one of those "LaTeX things". For
% instance, "\textbf{A} \textbf{B}" will typeset as "A B" not "AB". To get
% "AB" then you have to do: "\textbf{A}\textbf{B}"
% \thanks is no different in this regard, so shield the last } of each \thanks
% that ends a line with a % and do not let a space in before the next \thanks.
% Spaces after \IEEEmembership other than the last one are OK (and needed) as
% you are supposed to have spaces between the names. For what it is worth,
% this is a minor point as most people would not even notice if the said evil
% space somehow managed to creep in.

% The paper headers
\markboth{}%
{}
% The only time the second header will appear is for the odd numbered pages
% after the title page when using the twoside option.
% 
% *** Note that you probably will NOT want to include the author's ***
% *** name in the headers of peer review papers.                   ***
% You can use \ifCLASSOPTIONpeerreview for conditional compilation here if
% you desire.

% If you want to put a publisher's ID mark on the page you can do it like
% this:
%\IEEEpubid{0000--0000/00\$00.00~\copyright~2015 IEEE}
% Remember, if you use this you must call \IEEEpubidadjcol in the second
% column for its text to clear the IEEEpubid mark.

% use for special paper notices
%\IEEEspecialpapernotice{(Invited Paper)}

% make the title area
\maketitle

% As a general rule, do not put math, special symbols or citations
% in the abstract or keywords.
\begin{abstract}
This paper builds on top of a paper we have \RSCHANGE{published} very recently,
  in which we have proposed a novel approach to prime factorization
  (PF) by quantum annealing, where $8,219,999=32,749\times251$ was the highest prime product we were able to factorize ---which, to the
  best of our knowledge is the largest number which was ever
  factorized by means of a quantum device.
\JDst{The series of annealing experiments which led us to these results, however, did
not follow a straight-line path; rather, they involved a convoluted
trial-and-error process, full of failed or partially-failed attempts
and backtracks, which only in the end drove us to find the successful annealing
strategies.}  

In this paper, we delve into the reasoning behind our experimental
decisions and provide an account of some of the attempts we have
taken before conceiving the final strategies that allowed us to
achieve the results.  This involves also a bunch of ideas, techniques, and strategies we investigated which, although turned out to be inferior
wrt. those we adopted in the end, may instead provide insights to a
more-specialized audience of D-Wave users and practitioners.
\GMCHANGE{In particular, we show the following insights: ($i$) different initialization techniques affect performances, among which flux biases are effective when targeting locally-structured embeddings; ($ii$) chain strengths have a lower impact in locally-structured embeddings compared to problem relying on global embeddings; ($iii$) there is a trade-off between broken chain and excited CFAs, suggesting an incremental annealing offset remedy approach based on the modules instead of single qubits.}
Thus, by sharing the details of our experiences, we aim to provide insights into the evolving landscape
of quantum annealing, and help people access and effectively use D-Wave
quantum annealers.
\end{abstract}

% Note that keywords are not normally used for peerreview papers.

% For peer review papers, you can put extra information on the cover
% page as needed:
% \ifCLASSOPTIONpeerreview
% \begin{center} \bfseries EDICS Category: 3-BBND \end{center}
% \fi
%
% For peerreview papers, this IEEEtran command inserts a page break and
% creates the second title. It will be ignored for other modes.
\IEEEpeerreviewmaketitle

\section{Introduction}

{\em Quantum computing} has emerged as a
novel paradigm in computer science, offering the
potential capabilities to solve complex problems that have long
remained intractable for classical computers. Among the various
approaches within quantum computing, {\em quantum annealers (QA)} stand out
as a promising tool for tackling challenging computational tasks.
To this extent, {\em prime factorization (PF)} ---i.e., the problem of
breaking down a
number into its prime factors--- is a good candidate to be
effectively solved by quantum computing, in particular by quantum
annealing.
% While seemingly
% straightforward for small numbers, it becomes a difficult problem as
% the numbers involved grow larger.
This problem is of utmost
significance in modern cryptography, where the security of systems
often relies on the presumed computational intractability of
PF~\cite{rivest1978method}. 
Several approaches have been presented to address PF by quantum
computing,
e.g. \cite{Vandersypen2001,Lucero2012,MartinLopez2012,doi:10.1126/science.aad9480,PhysRevA.100.012305,selvarajan21},
by quantum annealing, e.g.  \cite{Dridi2017,Jiang2018,mengoni2020breaking},
or by hybrid quantum-classical technologies, e.g. \cite{Wang20,Karamlou21}. 
See \cite{math11194222,ding2023effective} for a summary.

This paper builds on top of 
a paper we have \RSCHANGE{published} very recently~\cite{ding2023effective}, in which we have proposed a
novel approach to PF by quantum annealing,
with two main results. 
%
%Our results are twofold.
First, we have presented a very compact modular {\em encoding} of a binary
multiplier circuit into the Pegasus QA architecture, which allowed us to
encode up to a 21$\times$12-bit multiplier (or alternatively a 22$\times$8-bit one)
into the Pegasus 5760-qubit topology of D-Wave Advantage
annealers.
Due to the modularity of the encoding, this number will scale up automatically with the growth of the qubit number in future chips. 
Second, we have investigated the problem of actually
\emph{solving} encoded PF problems by running an extensive
experimental evaluation on a D-Wave Advantage 4.1 quantum annealer.  In these
experiments we have introduced different approaches to initialize the
multiplier qubits, and adopted several performance-enhancement
annealing strategies.
Overall, within the limits of our QPU resources,
$8,219,999=32,749\times251$ was the highest prime product we were able to
factorize ---which, to the best of our knowledge, is the largest
number which was ever factorized by means of a ``pure'' quantum device
(i.e., without adopting hybrid quantum-classical techniques).

\JDst{The %series of
annealing experiments which led us to these results, however, did
not follow a straight-line path. Rather, they involved a convoluted
trial-and-error process, full of failed or partially-failed attempts
and backtracks, which in the end drove us to find the successful annealing
strategies.  
To this extent, if our research is an iceberg, then
Ding et al. (2024) %\cite{ding2023effective}
represents its visible part, finally 
emerging out of the water.} 

In this paper we \JDst{analyze the part of the iceberg which is under the
water level.
We} delve into the reasoning behind our experimental decisions and
provide a more comprehensive account of the steps and attempts we have
taken before conceiving the final strategies which allowed us to achieve the
results in \cite{ding2023effective}.
\JDst{In particular, w}We illustrate a bunch of ideas, techniques, and strategies we investigated
which, although turned out to be 
inferior wrt. those we  adopted in the end
---and as such were not of interest for the more general public
targeted in \cite{ding2023effective}---
may instead provide  insights to a more-specialized
audience of D-Wave QA users and
practitioners.
\GMCHANGE{In particular, we show the following insights:
  ($i$) different initialization techniques affect performance, among
  which flux biases are effective when targeting locally-structured
  embeddings; ($ii$) chain strengths have a lower impact in
  locally-structured embeddings compared to problems relying on global
  embeddings; ($iii$) there is a trade-off between a broken chain and
  excited CFAs, suggesting an incremental annealing offset remedy
  approach based on the modules instead of single qubits.} 
Thus, by sharing the details of our experiences, including both successes
and setbacks, we aim to provide insights into the evolving landscape
of quantum annealing and help people access and effectively use D-Wave
quantum annealers.

% \noindent
\JDst{Note to reviewers. As a consequence of what said above, we stress the fact that there is no overlapping between
 the scientific contribution of this paper and that of Ding et al (2024).}
% The very first letter is a 2 line initial drop letter followed
% by the rest of the first word in caps.
% 
% form to use if the first word consists of a single letter:
% \IEEEPARstart{A}{demo} file is ....
% 
% form to use if you need the single drop letter followed by
% normal text (unknown if ever used by the IEEE):
% \IEEEPARstart{A}{}demo file is ....
% 
% Some journals put the first two words in caps:
% \IEEEPARstart{T}{his demo} file is ....
% 
% Here we have the typical use of a "T" for an initial drop letter
% and "HIS" in caps to complete the first word.
\section{Methods}
%\begin{rschange}
We first summarize a few concepts from \cite{ding2023effective}.
The prime factorization problem (PF) of a biprime number $N$ can be addressed by SAT solvers by
encoding a $n\times m$ multiplier into a Boolean formula, fixing the
values of the output bits s.t. to encode $N$.
In \cite{ding2023effective}, we presented a modular {\em embedding} of a binary
multiplier circuit into the Pegasus QA architecture,
based on locally-structured embedding of SAT problems \cite{Bian2020}.
The multiplier circuit, represented in terms of a conjunction
of {\em Controlled Full-adder (CFA)} Boolean functions linked by means of
equivalences between variables,
is embedded  into the Pegasus topology,
with each CFA embedded into a 8-qubit tile
and with the variable equivalences %propagation of the two input variables to the multiplier
implemented through {\it chains}. % \JDCHANGE{(or {\it qubit sharing})}.
Each CFA $\Fx$ is encoded in terms of
a {\em penalty function}:
\begin{eqnarray}\label{eq:penfunction-ancillas}\textstyle
\Pxaz \defas \offset{} + \sum_{\substack{z_i\in V}} \theta_{i} z_i +
\sum_{\substack{(z_i,z_j)\in E, i<j}}
  \theta_{ij} z_i z_j;
%\ \ \ z_i \in \xs\cup\as;
  \ \ \ z_i \in \{-1, 1\};\\
  \label{eq:penfunction-ancillas2}
  \text{ } \ %\quad
  s.t.\     \forall \xs\ \ min_{\set{\as}} \Pxa
  \begin{cases}
= 0 &\text{ if }   F(\xs)=\top \\
   \geq g_{min} &\text{ if } F(\xs)=\bot
  \end{cases}
\end{eqnarray}
% ############## below copied from the first paper#############
where the Boolean variables \xs and \as are mapped into a subset
$\zs\subseteq V$ of the qubits in the topology graph $(V, E)$, 
s.t. the qubit
values $\set{1,-1}$ are interpreted as the truth
values $\set{\top,\bot}$ respectively;
$\theta_0$, $\theta_i$, $\theta_{ij}$ and $g_{min}$ are called respectively {\em
offset}, {\em biases},  {\em couplings} and the {\em gap};
the offset has no bounds, whereas the range for biases and couplings is
$[-4,+4]$ and $[-2,+1]$ respectively.
%##############################################################
(The ancilla variables $\as$ are needed %introduced 
to address the over-constrainedness of the encoding problem.)
The $\ts$ values in $\Pxa$ have been synthesized by means of 
\optimathsat{}~\cite{Sebastiani2020} s.t. to maximize $g_{min}$.%
\footnote{The bigger is $g_{min}$, the easier is for the annealer to
  discriminate solutions from non-solutions \cite{Bian2020}.}
% \JDCHANGE{
%   Notice that if the qubit sharing is applied in the embedding,
%   the bias of the shared qubit needs to be constrained
%   in the range of the qubit bias in the encoding.
% }
%(The larger $g_{min}$, the higher the probability to )
The penalty function of the whole multiplier is thus produced as the sum of
the penalty functions of the CFAs, plus a term ($2-2zz'$) for every chain
\tuple{z,z'}. Then it is fed to the annealer, forcing the values of the
output qubits so that to represent the biprime
number $N$, and forcing to $-1$ the value of the carry-in qubit of the rightmost CFA of
each row,
and the value of the in2 qubit of the CFAs in the first row in the multiplier. 
Therefore, if the annealer finds a ground state s.t. such penalty
function is zero, then the values of the qubit represent a solution of
the PF problem.
\footnote{\label{footnote:ancillas}From \eqref{eq:penfunction-ancillas2} we notice
  that, due to non-minimum values of $\as$, in principle we can have
  solution scenarios where $\Fx=\top$ and $0<\Pxa <
  g_{min}$, which we can %immediately
  recognize as solutions, or even s.t. $\Pxa\ge g_{min}$, for
  recognizing which we need testing $\Fx=\top$ explicitly, which can
  be performed very easily.%
% \JDCHANGE{\sout{Nevertheless, these scenarios never occurred in the
%     experiments in}} \cite{ding2023effective}.
}
(We refer the reader to \cite{ding2023effective} for a much
more detailed explanation.)

\subsection{Alternative approaches to initialize qubits}

Solving prime factorization of a specific number $N$ requires some of
the qubits to be initialized to some fixed value in \set{-1,1}. For
instance, given an 8-bit multiplier and $N=42$, whose binary representation is 
$00101010$,
then the qubits of the CFAs corresponding to the output number should be initialized respectively
to $\set{-1,-1,1,-1,1,-1,1,-1}$; also, e.g., the carry-in qubit of 
the CFA for the least significant bit in a number 
must be set to $-1$. D-Wave API offers a native function, ${\sf fix\_variables()}$,
that replaces the truth values of the qubits into the penalty
function. Unfortunately, this  causes a subsequent rescaling of all weights if one bias
or coupling does not fit into the proper range, reducing thus the gap
$g_{min}$ accordingly.

The initialization of qubits can be implemented either at the encoding
level (i.e. by imposing qubit values directly into the penalty
function \Px), or at the hardware level (i.e. by imposing the qubit
values through the tuning of the quantum annealer hardware). 
In \cite{ding2023effective} we adopted %tested
the latter implementation by
tuning flux biases, and showed the benefits they brought to the success probability of reaching the ground state. In this paper, we
mainly focus on the former type of implementation, proposing a few
alternatives to ${\sf fix\_variables()}$:

\begin{itemize}
    \item {\bf Ad-hoc encoding for the  CFAs}: we substitute the values of the input variables into the corresponding CFAs
and then re-encode these initialized CFAs, with reduced graphs,  into new CFA penalty functions. For instance, suppose we want to set the value of $c\_{in}$ to false. Then we feed to the OMT solver the
          extended formula
          $F'(\xs) = F(\xs) \wedge \neg c\_{in}$ to
          generate a new specialized
          penalty function.
          % \JDSIDENOTE{Why next sentence? ``To... CFAs.'' I'd drop it.}
          % \JDSIDENOTE{Answer: The reason is that due to the qubit sharing,
          % if these special CFAs are encoded independently,
          % the minimal gap can not be guaranteed within the range.}
          To prevent the $g_{min}$ from being scaled down
due to the input values, during the re-encoding process we take into
account all combinations of possible inputs that occur in the CFAs.
\footnote{This is made necessary by one further technique,
    namely {\em qubit
    sharing}, which  we have introduced
  in \cite{ding2023effective} and which is not explained here.}
This results into the generation of an {\it offline library of specialized CFAs},
%\JDCHANGE{\sout{The re-encoding process produces penalty functions}}
with increased minimal gaps, $g_{min} \in [3, 18]$.
Notice that, using these modified encodings, we obtained some solutions where
 $\Fx=\top$ and $0<\Pxa <g_{min}$ (recall Footnote~\ref{footnote:ancillas}),
which never occurred in the experiments reported in
\cite{ding2023effective}.
%% Advertising these facts is not among te goal of this paper
Both the gap increment and the extra solutions
can increase the probability
to find solutions.

% \JDTODO{the next sentence is meaningless. ??}
% \JDTODO{Answer: the following phenomena did not occur in the penalty function
% of the generic CFA. But it can increase the success probability.}
%
% \ignore{Moreover, given a  satisfying assignment $\xs$, it occurs that $P_F(\as|\xs\models F) < g_{min}$ for some $\as$, further increasing the success probability
% in finding solutions for PF problems.}
% \TODO{Why is next sentence here? Haven't we said it already in
%   \cite{ding2023effective}? Shall we put it in the preamble part
%   of this section? Shall we drop it?}

%, further increasing the probability of success.}

\item {\bf Extra chaining}:
%\RSCHANGE{
    we notice that in the penalty functions 
    of CFA we have obtained, the biases of the qubits are all
    within  $[-1, 1]$,
    whereas the range for the D-Wave Advantage 4.1 is $[-4, 4]$.
    Based on these facts, we have explored a simple alternative way to initialize qubits,
    without the risk of rescaling down the $g_{min}$ value.
    Specially, in order to assign qubit $z$ to $1$ [resp. $-1$], we can 
%    \JDCHANGE{follow the definition of the penalty function to}
    add the penalty function for $z=1$ $(2-2z)$ [resp. for $z=-1$ $(2+2z)$] to the penalty function of the multiplier, 
    s.t. the bias of $z$ safely remains in $[-3,3]$.
    Equivalently, we can find an unused neighbour qubit $z'$ (if any), 
    add an equivalence chain between $z$ and $z'$,
%    resulting in a new penalty function extended with $2-2zz'$; 
    and then initialize $z'$ to
    $1$ (resp $-1$) by ${\sf fix\_variables()}$.
\end{itemize}

% \JDSIDENOTE{I prefer to remove this section,
% since the number of reduced variables
% hadn't been verified (I mean there is no test for this results)}{\subsection{Exploiting Boolean Value Propagation}}
    \JDst{In the experiments, we compare these initialization approaches
        for D-Wave Advantage systems on factoring small integers (of up to $5\times5$ bits),
        with the annealing time set to $10\mu s$ and 1000 samples for each problem instance.}
% We can produce a simpler penalty 
% function \Pxa{} of our multiplier by simplifying upfront the input Boolean formula $\Fx$.
%   As we have noticed above, some qubits must be initialized to
%   some fixed values.
%   % , corresponding to Boolean values of variables in
%   % the Boolean formula $F(\allx)$ representing the multiplier.
%   It may be the case, however, that
%   some such values force the value of some other Boolean variables in $F(\allx)$. (E.g., consider
%   $F(\allx)\defas ... \wedge (x_1\iff (x_2 \vee x_3))$: if $x_1$ is
%   forced to be $-1$, then both $x_2$ and $x_3$ are forced to be $-1$; if
%   instead $x_2$ is forced to be $1$, then $x_1$ is forced to be
%   $1$.)
% %
%   Thus, we can detect these derived values by applying {\it
%     Boolean value propagation (BVP)} to \Fx. Then we set the
%   values of  the corresponding qubits by calling
%   ${\sf fix\_variables()}$. This allows us to further reduce the number of binary
%   variables for the annealer to cope with.
% %
% Notice that, since modern Boolean reasoning tools 
% are capable of performing BVP very efficiently, the overhead of performing it as a preprocessing step 
% is negligible.
% The effects of BVP are reported in Table
% \ref{tab: instantiations} and further discussed in Section 3.

\subsection*{\JDst{2.2 Experimenting with different chain strengths}}

\JDst{In the encoding of our multiplier, we always used the strongest coupling strength, i.e., $c=2$,
for all chains. 
In order to verify that this is the best choice, 
we compare different chain strengths, i.e. $c\in\{2, 1.5, 1\}$, on factoring integers of up to 8$\times$8 bits,}
%\JDCHANGE{
%    whose size is hard for the system to find a solution
    \JDst{with the annealing time set to $10\mu s$ and 1000 samples.
        This is implemented by changing the penalty functions of every chain
        from $2-2zz'$ to $c-czz'$.}
    % The implementation of these different chain strengths
    % correspond to the changes of the penalty functions of chains,
    % i.e., from $2-2zz'$ to $c-czz'$.
%}
%  This modification only required us to update the penalty function of the multiplier, 
%  modifying the couplings of the chained qubits accordingly.

\subsection{The impact of chain strength in QA for modular encodings}

\GMCHANGE{
 The effect of chain strength has been previously studied 
    in the context of {\em global embedding}, where
    the input problem is first encoded into a QUBO problem, which is then embedded into the hardware by means of embedding algorithms. The main issue of that approach is that the QUBO model does not know in advance how many chains are there in a specific topology and where they will be placed. Thus, the addition of chains a-posteriori---whose length and placement are out of the control of the user---and the consequent re-scaling of biases and coupling may affect the performances of the algorithm.
}

\GMCHANGE{
    Our {\em locally-structured embedding} approach in
    \cite{ding2023effective} differs from the above approach because the Ising model that is generated is already hardware-compliant, so that there is no risk of weights rescaling, and we do not need a fine-grained analysis of chain strength. Given the modularity of our encoding and the presence of chains to allow communication between neighboring modules, however, it is still important to investigate the side effects of chain strength in modular encoding.
    To this extent, we choose
    a set of chain strengths, $c\in\{1, 1.5, 2\}$,
    as representatives for
    investigating their effects on the performance of our locally structured embedding approach on QA systems. 
}

\subsection*{\JDst{2.3 Analyzing excitations of chains and CFAs}}
\JDst{     We analyze the performances of the QA to solve the factorization
     of $8\times8$-bit problems,
     with the annealing time set to $10\mu s$ and 1000 samples. 
     In particular, 
    we obtain two statistics %from all the 10 problem instances 
    about ($i$) the number of broken occurrences of each
    chain ---i.e., when the qubits connected by the chain have opposite values--- out of 1000 samples; 
    ($ii$) the number of 
      excitation
      occurrences of a CFA
    ---i.e. when 
      the value of
      the CFA penalty function is higher than $g_{min}$--- out of 1000 samples.}
 
\jdREVISION{
    \subsection{Incrementally remedying excited CFAs}
    % By further examining the behaviors of chains and CFAs at specific coordinates
    % for the small sizes tested in Section 2.2, 
    % we find in Figure \ref{fig: diff_chain_strengths}-2)
    % that besides the consistency with the general behavior 
    % shown in Figure \ref{fig: diff_chain_strengths}-1),
    % the number of occurrences of CFAs getting excited
    % is non-homogeneous.
}
\JDst{
    From the statistics about broken chains and excited CFAs obtained in Section 2.3,
    we observed that the probability for CFAs
      to reach the ground state are non-homogeneous.    
}
We assume that if all CFAs in a multiplier reach the ground state
with high probability, then the success probability of the whole
multiplier will be positively affected. Based on this assumption,
we have {proposed} an incremental remedy strategy, to remedy the most excited
CFAs during the solving process.

{The remedy approach} is based on {\it anneal offsets} \cite{QPUannealing}.
In the standard annealing process of D-Wave systems,
the annealing schedules are set identically for all qubits.
However, the system also allows for adjusting the annealing schedule for each qubit.
This is implemented by offsetting the global, time-dependent bias signal $c(s)$ 
that controls the annealing process.
More specially, for a qubit $q_i$, 
its anneal offsets $\pm \delta c_i\neq 0$ correspond to 
advancing and delaying the standard annealing schedule, respectively.

\GMCHANGE{
    In a fashion similar to \cite{extraA, extraB, extraC, extraD} we adopted the idea of incrementally fixing annealing offset weights to increase the probability of reaching a ground state. Differently from these papers, however, where the annealing offset is set to qubits, we set modules of our encoding (i.e. CFAs) as the target of annealing offset tuning, and we choose the number of excitations of these modules
    as a measure to guide the remedy strategy process.
}

In each step of our incremental remedy approach, 
we first find the most-excited CFA
---i.e. the CFA whose number of excitation occurrences out of the 1000 samples
is maximum---
and then continue to advance the annealing process of all its qubits by annealing offset $\delta c_i=0.01$,
on top of the previous remedying history,
until the CFA is no longer the most excited.
The procedure terminates either if the system reaches one ground state or if it reaches 
\JDst{a certain number of iterations.
In our experiments, due to the limited access of QPU time, the upper bound of iterations has been set as the number of bits of the output number. 
We tested this strategy, together with flux biases instantiation, to factor $8\times8$ and $9\times9$ bits integers. 
} \GMCHANGE{a certain number of steps set as a threshold.
This threshold is chosen according to the limitation on the access of QuPU, e.g., the perimeter of the multiplier embedded.}

\section{Results}

%

%\JDTODO{Here, for each approach, refer explicitly to Table/figure number.}

\subsection{Results of different initialization approaches of qubits}

In the experiments, we compare the proposed initialization approaches
    on D-Wave Advantage system 4.1 
    for factoring small integers of up to $5\times5$ bits,
    with the annealing time ($T_a=10\mu s$) and 1,000 samples 
    set for each problem instance.
\JDst{An extensive comparison of the initialization approaches, 
addressing small prime factorization problems is shown in}
Table~\ref{tab: instantiations}.

\begin{table}[!t]
    \centering
    % \begin{minipage}{.15\textwidth}
    %     % \begin{figure}
    %         \centering
    %         \includegraphics[width=1\linewidth]{figs/multiplier_ver2_CFA0/CFA0_embedding_big.png}
    %         \caption{\label{fig: CFA0} The ising model of CFA0 of $g_min=2$.}
    %     % \end{figure}
    % \end{minipage}
    % \begin{minipage}{.3\textwidth}
    %     % \begin{table}
    %     \centering
    %     \scriptsize
    %     \begin{tabular}{|l|r|r|}
    %         \toprule
    %         \multirow{2}{*}{size}    & \multirow{2}{*}{inputs} & \multirow{2}{*}{CFA0}  \\
    %         & & BVP + api \\
    %         \midrule
    %         \multirow{3}{*}{3$\times$3}
    %         & 25   & 68     \\           
    %         & 35   & 244    \\           
    %         & 49   & 224    \\           
    %         \midrule
    %         \multirow{3}{*}{4$\times$4}
    %         & 121  & 34     \\             
    %         & 143  & 45     \\             
    %         & 169  & 48     \\             
    %         \midrule
    %         \multirow{15}{*}{5$\times$5}
    %         & 289=17$\times$17   & 3     \\
    %         & 323=17$\times$19   & 2     \\
    %         & 361=19$\times$19   & 8	 \\  
    %         & 391=17$\times$23   & 48    \\
    %         & 437=19$\times$23   & 54    \\
    %         & 493=17$\times$29   & 38    \\
    %         & 527=17$\times$31   & 63    \\
    %         & 529=23$\times$23   & 2	 \\  
    %         & 551=19$\times$29   & 25    \\
    %         & 589=19$\times$31   & 34    \\
    %         & 667=23$\times$29   & 19    \\
    %         & 713=23$\times$31   & 3	 \\  
    %         & 841=29$\times$29   & 15    \\
    %         & 899=29$\times$31   & 26    \\
    %         & 961=31$\times$31   & 4	 \\  
    %         \midrule
    %     \end{tabular}
    % \end{minipage}
    \begin{minipage}{1\textwidth}
        % \begin{table}
            \centering
            \scriptsize
            \begin{tabular}{|r|r|rrrr|rrr|}
                \toprule
                \multirow{2}{*}{size}    & \multirow{2}{*}{inputs} &\multicolumn{4}{c}{CFA0}                & \multicolumn{3}{c}{CFA1}\\
                                         &                         & api & ad-hoc & chain & flux-bias       & api & chain & flux-bias\\
               \midrule
               \multirow{3}{*}{3$\times$3}
                & 25=5$\times$5    & 161  & 154 & 93  & 308  	    & 327  & 173 & 136   \\
                & 35=5$\times$7    & 389  & 666 & 286 & 711  	    & 410  & 379 & 951   \\
                & 49=7$\times$7    & 450  & 577 & 312 & 906  	    & 344  & 295 & 997   \\
                \midrule
                \multirow{3}{*}{4$\times$4}
                & 121=11$\times$11   & 17   & 4   & 30  & 63       & 9    & 33  & 0     \\
                & 143=11$\times$13   & 40   & 52  & 28  & 129	    & 122  & 32  & 67    \\
                & 169=13$\times$13   & 31   & 54  & 4	 & 312      & 84   & 69  & 5     \\
                \midrule
                \multirow{15}{*}{5$\times$5}
                & 289=17$\times$17     & 5    & 0   & 0  &	1  		& 3    & 1   & 0     \\
                & 323=17$\times$19     & 2    & 0   & 1  &	7  		& 22   & 3   & 0     \\
                & 361=19$\times$19     & 1    & 1   & 0  &	1  		& 11   & 1   & 3     \\
                & 391=17$\times$23     & 6    & 1   & 4  &	119 	& 5    & 19  & 9     \\
                & 437=19$\times$23     & 17   & 0   & 3  &	67  	& 3    & 2   & 0     \\
                & 493=17$\times$29     & 3    & 6   & 0  &	4  		& 8    & 0   & 2     \\
                & 527=17$\times$31     & 21   & 11  & 6  &	91  	& 6    & 5   & 37    \\
                & 529=23$\times$23     & 5    & 0   & 3  &	8  		& 0    & 1   & 8     \\
                & 551=19$\times$29     & 0    & 11  & 4  &	24  	& 2    & 3   & 4     \\
                & 589=19$\times$31     & 16   & 13  & 11 &	7  		& 1    & 22  & 52    \\
                & 667=23$\times$29     & 0    & 6   & 2  &	3  		& 8    & 9   & 105   \\
                & 713=23$\times$31     & 11   & 12  & 3  &	26  	& 2    & 1   & 138   \\
                & 841=29$\times$29     & 5    & 9   & 8  &	148 	& 14   & 8   & 7     \\
                & 899=29$\times$31     & 17   & 76  & 5  &	222 	& 7    & 13  & 343   \\
                & 961=31$\times$31     & 1    & 43  & 0  &	37  	& 1    & 0   & 338   \\
                \midrule
            \end{tabular}
        % \end{table}
    \end{minipage}
    \caption{\label{tab: instantiations} Different initialization approaches
            for solving small-size PF, with the annealing time $T_a=10\mu s$
            and 1,000 samples for each problem instance.}
\end{table}

By comparing the performances of the initialization techniques
\jdREVISION{shown in Table \ref{tab: instantiations}}, 
we notice that the ad-hoc re-encoding outperforms the native API and the extra-chain approaches,
but it still does not perform as well as the flux-bias tuning, which
we finally adopted in \cite{ding2023effective}.
In \cite{ding2023effective} we also proposed a variant of the CFA function, namely CFA1, 
minimizing the number of unsatisfying assignments with $g_{min}$ equal to 2. 
For the sake of completeness, we also tested this encoding in combination with initialization techniques other than flux biases. 
These results confirm that the combination of the flux-bias initialization 
and the improved CFA1, which we adopted in \cite{ding2023effective}, produces the highest success probability
for D-Wave Advantage 4.1 in finding solutions.
%\JDCHANGE{
For this reason, we continue to use this combination, 
the flux-bias initialization + CFA1,
in the following experiments of this paper.
%}

% \subsection{Exploiting Boolean Value Propagation}

% The results shows some minor improvements achieved by incorporating
% BVP as a pre-processing step. We highlight, however, that the results
% are quite below the strategy we finally adopted in
% \cite{ding2023effective}. Even testing lower or higher annealing times
% $T_a\in \{1, 10, 20\}\mu s$ do not benefits BVP.
 
% We also empirically noticed how the number of propagated values only
% grows linearly with respect to the input number size, thus the effects
% of BVP are less evident with a prime factorization of larger numbers.
% \ignore{We empirically noticed that although the number of propagated
%   values grows only linearly with the bit-size of the factors, we have
%   a significant increase in the number of solutions the annealer can
%   find.}

\subsection*{\JDst{3.2 Results on different chain strengths}}

% \setcounter{figure}{0}
% \begin{figure}
%     \includegraphics[width=1\textwidth]{figures/Figura1.jpg}   
%     \caption{\label{fig: extrafigure} Results and statistics on small and middle-size PF problems. \textbf{(1st and 2nd rows).} 
%       %\JDCHANGE{[Via flux-bias initialization + CFA1]}
%       The impact of different chain strengths $\in \{-2, -1.5, -1\}$,
%     for solving small PF problems, 
%         with the annealing time $T_a=10\mu s$
%         and 1,000 samples for each problem instance.
%         \textbf{(3rd and 4th rows).} Cumulative results of the excited chains (3rd row) and CFA1 (4th row)
%     for factoring 10 integers of $8+8$ bits, 
%         with the annealing time $T_a=10\mu s$
%         and 1,000 samples for each problem instance.}
% \end{figure}

\JDst{In Figure (1st row) we report the
number of samples in which: ($i$) the qubits of a chain are equivalent
for all the chains (dotted lines);
and ($ii$) the Ising model of all CFAs  
behaves correctly (solid line).
In Figure (2nd row)  
we plot the number of samples reaching the ground state.
The results show that the strongest chain coupling did not always
produce the highest success probability for small factorization problems (such as the $6\times6$ and $7\times7$ problems), as per our assumptions.
However, the stronger chain couplings indeed succeeded 
in finding the prime factors to one 16-bit integer, 59,989.}

    % \newpage
    \subsection{Results of different chain strengths}

    \begin{figure}[t]
        \centering
        \begin{minipage}{.46\linewidth}
            \includegraphics[width=1\linewidth]{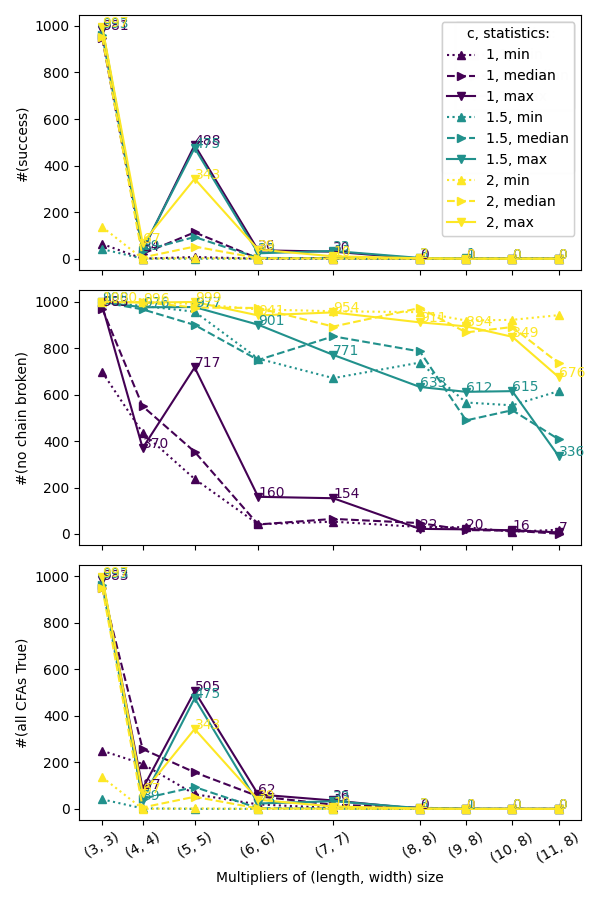}
        \end{minipage}
        \begin{minipage}{.52\linewidth}
            \includegraphics[width=1\linewidth]{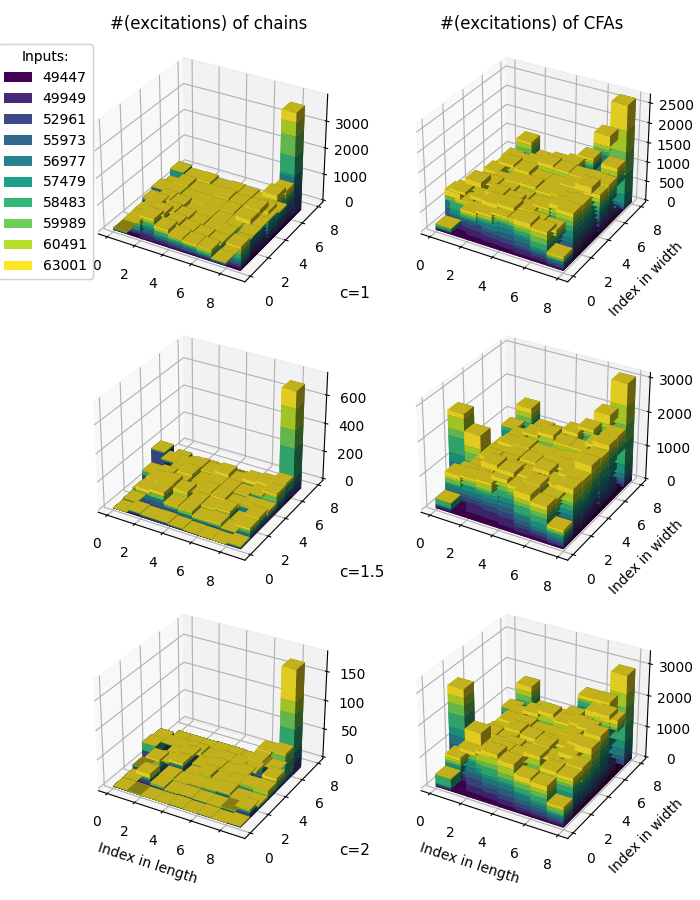}
        \end{minipage}
        \caption{
            \GMCHANGE{
            \label{fig: diff_chain_strengths}
            (Left) Comparison of different chain strengths $c, c \in \{1, 1.5, 2\}$, for 
            QA to factor integers of $3+3$ bits up to $11+8$ bits,
            with the annealing time $T_a=10\mu s$ and 1,000 samples 
            set for each problem instance. \\
            \label{fig: excitations}
            (Right) Excitations distribution of chains (first column) and CFAs (second column)
            for factoring 10 integers of $8+8$ bits tested in the previous experiments,
            with chain strength equal to $c\in\{1, 1.5, 2\}$ (respectively top, middle, and bottom row).
        }
        }
    \end{figure}

    \GMCHANGE{Using the initialization approach based on CFA1 + flux biases and the same configuration of the annealing system
    ($T_a=10\mu s, 1,000$ samples for each problem instance) of previous experiments, 
    we test different chain strengths ($c \in \{1, 1.5, 2\}$)  
    for QA factoring integers from $3\times 3$ up to $11\times 8$ bits, using the 10 highest co-prime number for each multiplier size.}

    \GMCHANGE{
    In Figure \ref{fig: diff_chain_strengths} (left) we summarize the
    results of all samples provided the QA. Sorting them by the
    size of the input problem ($x$-axis), we plot respectively the number of samples successfully reaching the ground state (first plot), the number of samples having no broken chain (second plot), and the number of samples having no excited CFA (third plot). In general, we report the score of the median sample among all problems (dashed line) as a summary of the annealer behavior for each sample size. In addition, for each sample size, we provide information on the problem that reaches the ground state the least frequently (represented by the minimum dotted line in Figure \ref{fig: diff_chain_strengths} (left)), as well as the one that reaches the ground state the most frequently (represented by the maximum solid line in Figure \ref{fig: diff_chain_strengths} (left)). }
    
    \GMCHANGE{
    We see that
    stronger chains ($c \in \{1.5, 2\}$) do not always
    bring us a higher success probability in general for the chosen problem sizes,
    and that weaker chains ($c=1$) can produce higher success probabilities
    than stronger chains occasionally for middle-size problems.
    Notice that this result, in terms of the success probability,
    is consistent with what is mentioned by \cite{extraA}, suggesting that locally-structured embedding does not behave differently from global embedding regarding chain strengths.
    We also observe that as the problem size increases,
    weaker chains tend to be broken more easily than stronger chains.
    The rapidly declining dotted yellow lines confirm this phenomenon,
    approaching $0$ for problems of bigger size.
    Based on these two observations, we speculate that the strongest chain, which was chosen in \cite{ding2023effective},
    is the best candidate for factoring integers of up to $17\times8$ bits,
    the maximal problem size they could encode into the target QA system
    with a locally-structured embedding.}

    % To close this gap seems able to increase the success probability.
    % Based on this assumption, we proposed an incremental remedy
    % for excited (equivalently, functioning incorrectly) CFAs to close this gap.

\subsection*{\JDst{3.3 Results on excitations of chains and CFAs}}

\JDst{The results on excitations of chains and CFAs are reported as 3D bar plots in Figure (3rd and 4th row, respectively). Each problem instance is mapped with its color. The $x$ and the $y$ axis
    correspond to the column and row of the multiplier respectively;
    the $z$ axis represents the sum of excitations of each chain or CFA for the tested ten problem instances.
}

\JDst{    This local analysis highlights two main behaviors of our multipliers. First, there is a trade-off between broken chains and the excitations of CFAs: 
the weaker the chains are, the more likely they are broken,
  and the fewer the samples where the CFAs is excited. Second, the excitations of CFAs are not uniformly distributed.}

\subsection*{\JDst{3.4 Results on incrementally fixing excited CFAs}}

\JDst{The results of our fixing strategy are presented in Table.
The presence of some $0$ in Table demonstrates the partial effectiveness of this local fixing,
when the required annealing time is reaching the limit.
Note that the effective annealing time is assumed to be confined, 
e.g., $\lesssim 20\mu s$, 
that increasing it would not bring significant improvement
on the success probability of finding solutions.
Therefore, this technique can be used as a temporary approach 
for the system to solve slightly larger problems 
before the effective annealing time of systems is improved.
For the sake of completeness, we also report the detailed fixing history for the factorization of $113,507$ in Figure.}

% \setcounter{figure}{2}
% \setcounter{subfigure}{0}
% \begin{subfigure}
%     \includegraphics[width=\textwidth]{figures/Figura2.jpg}
%     \setcounter{figure}{2}
%     \setcounter{subfigure}{-1}
%     \caption{\label{fig: local_fixing_for_factoring_113507} Fixing results
%     for factoring 16-bit 113,507 
%         in Table~\ref{tab: local_fixing}.
%         Each coordinate corresponds to the CFA in the multiplier circuit.
%         The numbers in the figures in the leftmost and rightmost columns represent
%         the number of excitations out of 1000 samples,
%         whereas the number in the figure in the middle denotes
%         the anneal offset used in the whole fixing process,
%         for advancing the annealing schedule of the specific CFA.
%     % \sout{\textbf{Left:} Frequency of each CFA
%     % of being excited, out of 1000 samples, for the first
%     % iteration. \textbf{Middle:} The distribution of the annealing
%     % offset scores for the last iteration. 
%     % \textbf{Right}: Frequency of each CFA of being excited, out of
%     % 1000 samples, for the last iteration. }
%     Notice the less homogeneous
%     distribution of excitations of CFAs in the leftmost figure compared to the rightmost figure. \textbf{(Left).} Before the fixing. \textbf{(Middle).} The fixing history. \textbf{(Right).} After the fixing.
%     }
% \end{subfigure}

    % \newpage
    \subsection{Results of incrementally fixing excited CFAs}

    \GMCHANGE{From the experiment of the previous subsection, we can see that there seems to be a trade-off between broken chains and the excitations of CFAs: 
the weaker the chains are, the more likely they are broken,
  and the fewer the samples where the CFAs are excited. Moreover, the excitations of CFAs are not uniformly distributed. To this extent, we studied the distribution of broken chains and CFAs in 10 $8\times8$ factoring problems, shown in Figure \ref{fig: diff_chain_strengths} (right). The results on excitations of chains and CFAs are reported as 3D bar plots in Figure (3rd and 4th row, respectively). Each problem instance is mapped with its color. The $x$ and the $y$ axis
    correspond to the column and row of the multiplier respectively;
    the $z$ axis represents the sum of excitations of each chain or CFA for the tested 10 problem instances. These results support testing an incremental remedy strategy based on modules.}

  \begin{figure}[t!]
    \centering
    \begin{subfigure}[b]{\textwidth}
        \centering
        \includegraphics[width=\textwidth]{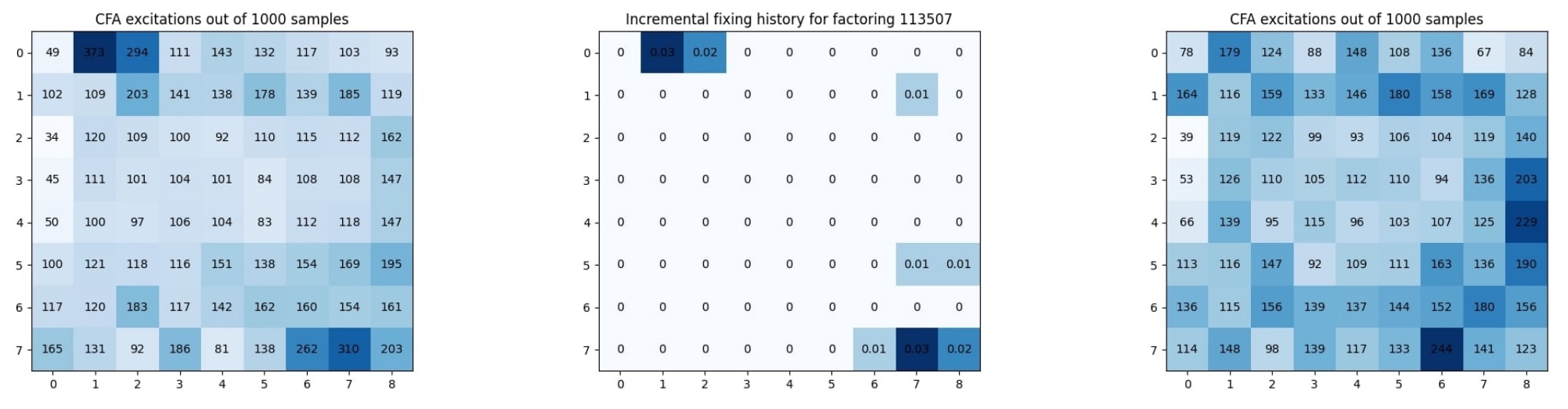}
        
    \end{subfigure}
    \caption{
            Fixing results for factoring 16-bit 113,507 in Table~\ref{tab: local_fixing}.
            Each coordinate corresponds to the CFA in the multiplier circuit.
            The numbers in the figures in the leftmost and rightmost columns represent
            the number of excitations out of 1000 samples,
            whereas the number in the figure in the middle denotes
            the anneal offset used in the whole fixing process,
            for advancing the annealing schedule of the specific CFA.
            Notice the less homogeneous distribution of excitations of CFAs in the leftmost figure compared to the rightmost figure. 
            \textbf{(Left).} Before the fixing. 
            \textbf{(Middle).} The fixing history. 
            \textbf{(Right).} After the fixing.
        }
        \label{fig:local_fixing_for_factoring_113507}
\end{figure}

    \GMCHANGE{With the strongest chain strength and the same configuration of the annealing system as the other experiments in the paper,
    we test the approach of incrementally fixing excited CFAs
    for QA factoring the highest integers 
    of $8\times8$ bits up to $10\times8$ bits from the experiments shown in Figure \ref{fig: diff_chain_strengths}. The results are shown in Table \ref{tab: local_fixing}. To get an extensive analysis of the novel remedy strategy, we tested three different configurations, with the only difference being the number of samples obtained for each iteration (respectively 1000, 2000, and 3000). We also show in Figure~\ref{fig:local_fixing_for_factoring_113507} the behavior of the remedy strategy on one of the problem instances.}

    From the results, we can see that the remedy strategy helps in solving some of the problem instances. In particular, this approach works under the assumption the user has a limited amount of QPU time (i.e. the annealing time is confined to values $\lesssim 20\mu) s$, showing its effectiveness when users are bound to tight constraints in accessing the D-Wave devices. This approach works more effectively with smaller instances, reaching the ground state more frequently and with fewer iteration steps. Moreover, increasing the sample size does not impact performances, showing sporadically improvements in reaching the ground state when the number of samples increases. Nevertheless, setting the annealing offset scores based on the modules' properties instead of targeting qubits independently seems promising, and further investigations could define different conditions to prioritize the annealing of some CFAs.

\section{Discussion}

This paper has built upon the recent work presented in our previous publication~\cite{ding2023effective}, which introduced a novel approach to the problem of PF through quantum annealing. In contrast to our previous paper, which showcased exclusively the effective techniques that highly benefited our task, here we discussed several intermediate and less successful approaches. This comprehensive exploration provides insights into the intricacies that influenced our final results in~\cite{ding2023effective}. The code to replicate these experiments is reported in the following publicly available repository: \url{https://gitlab.com/jingwen.ding/multiplier-encoder-2nd}.

Our experiments revealed several insights:

\begin{itemize}
    \item \textbf{Effectiveness of Flux Biases Tuning:} We showed that
      the techniques to initialize qubits implemented at the encoding
      level were not as effective as flux-biases
      tuning. Nevertheless, they can be considered as viable
      alternative to the usage of {\sf fix\_variables()} in other contexts.
    \item \textbf{Chains Coupling Strength:} Even though using the
      highest value for chains coupling strength might not be optimal
      for small-sized problems, it proved crucial for solving more
      complex problems. This highlights the delicate balance between
      problem size and annealing parameters, e.g. chain strength.
    \item \textbf{Trade-off Between Broken Chains and CFA
        Excitations:} We observed a trade-off between the presence of
      broken chains and the excitations of CFAs when the QA generates
      its samples. This further highlights the importance of
      monitoring chain strength in other contexts.
    \item {\bf Non-Uniform Distribution of CFA Excitations:} The
      excitations of CFAs were found to be non-uniformly distributed
      for different samples on the same problem
      instance. Understanding this distribution can be valuable for
      tailoring annealing strategies to specific problem instances. 
    \item {\bf Remedy Strategy for Middle-Size Problems:} The remedy
      strategy we proposed in Section 2.4, based on the above
      observations, showed minor benefits in solving middle-sized
      problems. Nevertheless, it could be useful in other contexts.
\end{itemize}

By delving into the details of our experimental journey, listing both our successes and setbacks, we aim to provide valuable insights to a more specialized audience of D-Wave Quantum Annealer users and practitioners. Our work contributes to the evolving world of quantum annealing and equips researchers and professionals with additional knowledge to effectively use D-Wave quantum annealers in their applications.

% if have a single appendix:
%\appendix[Proof of the Zonklar Equations]
% or
%\appendix  % for no appendix heading
% do not use \section anymore after \appendix, only \section*
% is possibly needed

% use appendices with more than one appendix
% then use \section to start each appendix
% you must declare a \section before using any
% \subsection or using \label (\appendices by itself
% starts a section numbered zero.)
%

\newpage
\begin{sidewaystable}[t!]
  \caption{\label{tab: local_fixing} %\JDCHANGE{[Via flux-bias initialization + CFA1]} 
  Results of incrementally remedying excited CFAs
  for factoring integers of $8\times8$ bits up to $10\times8$ bits,
  with the same annealing time $T_a=10\mu s$
  with the number of samples ranging
  from 1,000 to 3,000 set for each problem instance.
  For each problem, we first report the starting point sample, including its energy, the most excited CFA, and the number of its excitations respectively. Then, we report the number of iterations performed by the remedy strategy (a bold number means we did not reach the step threshold and a ground state has been found), together with the energy and the current most excited CFA. 
  }
  \centering
  % \scriptsize
  \footnotesize
  \setlength{\tabcolsep}{4pt}
  \begin{tabular}{|l|c||rl|rrl||rl|rrl||rl|rrl|}
      \toprule
      \multirow{2}{*}{size}    & \#samples  & \multicolumn{5}{c}{1000}   & \multicolumn{5}{c}{2000} & \multicolumn{5}{c}{3000}\\
                               & input      & $P_F$ & (CFA, \#excs) & i & $P_F$ & (CFA, \#excs)
                                            & $P_F$ & (CFA, \#excs) & i & $P_F$ & (CFA, \#excs)
                                            & $P_F$ & (CFA, \#excs) & i & $P_F$ & (CFA, \#excs) \\
      \midrule
      \multirow{10}{*}{8$\times$8}
      & 49447=251$\times$197     & 6.25	  & ((6, 7), 395)	   & 5	  & \textbf{0.0}		& ((0, 1), 315)       & 6.0	    & ((0, 1), 1268)& 32	& 4.083		& ((6, 5), 490)         & 4.0	    & ((0, 1), 1377)	& 32	& 4.083		& ((4, 7), 894)\\
      & 49949=251$\times$199     & 6.083	  & ((6, 7), 466)	   & 4	  & \textbf{0.0}		& ((7, 6), 347)       & 4.0	    & ((0, 1), 982)	& 32	& 4.0		& ((6, 5), 429)         & 2.0	    & ((0, 1), 1461)	& 25	& \textbf{0.0}		& ((5, 7), 631)\\
      & 52961=251$\times$211     & 2.083	  & ((7, 5), 454)	   & 32	  & 6.083	& ((6, 7), 250)       & 6.167	 	& ((7, 5), 994)	& 32	& 6.167		& ((5, 7), 452)         & 6.083		& ((7, 5), 1705)	& 6	    & \textbf{0.0}		& ((7, 6), 1001)\\
      & 55973=251$\times$223     & 4.0	  & ((5, 7), 242)	   & 2	  & \textbf{0.0}		& ((7, 3), 267)       & \textbf{0.0}	    & ((0, 1), 569)	& 0	    & \textbf{0.0}		& ((0, 1), 569)         & \textbf{0.0}	    & ((7, 6), 896)	    & 0	    & \textbf{0.0}		& ((7, 6), 896)\\
      & 56977=251$\times$227     & 2.083	  & ((7, 6), 457)	   & 31	  & \textbf{0.0}		& ((0, 1), 351)       & 4.083	 	& ((7, 6), 921)	& 8	    & \textbf{0.0}		& ((7, 6), 593)         & 4.083		& ((7, 6), 1555)	& 16	& \textbf{0.0}		& ((0, 1), 739)\\
      & 57479=251$\times$229     & 6.0	  & ((5, 7), 277)	   & 32	  & 4.0		& ((6, 5), 200)       & 4.083	 	& ((0, 1), 722)	& 32	& 4.083		& ((6, 6), 471)         & 4.0	    & ((5, 7), 925)	    & 1	    & \textbf{0.0}		& ((0, 1), 905)\\
      & 58483=251$\times$233     & 4.083	  & ((7, 7), 338)	   & 32	  & 4.0		& ((0, 3), 242)       & 4.083	 	& ((7, 7), 779)	& 32	& 4.083		& ((1, 4), 452)         & 4.0	    & ((7, 7), 1069)	& 32	& 4.0		& ((2, 1), 669)\\
      & 59989=251$\times$239     & \textbf{0.0}	  & ((7, 7), 252)	   & 0	  & \textbf{0.0}		& ((7, 7), 252)       & \textbf{0.0}	 	& ((7, 7), 815)	& 0	    & \textbf{0.0}		& ((7, 7), 815)         & \textbf{0.0}		& ((7, 7), 1282)	& 0	    & \textbf{0.0}		& ((7, 7), 1282)\\
      & 60491=251$\times$241     & 2.0	  & ((7, 7), 237)	   & 32	  & 4.083	& ((7, 7), 276)       & 2.0	    & ((7, 7), 856)	& 32	& 2.0	    & ((1, 4), 461)         & 2.0	    & ((7, 7), 1082)	& 32	& 2.0		& ((3, 0), 589)\\
      & 63001=251$\times$251     & 4.083	  & ((7, 7), 492)	   & 4	  & \textbf{0.0}		& ((0, 2), 292)       & 2.0	    & ((7, 7), 836)	& 1	    & \textbf{0.0}		& ((7, 7), 889)         & 2.0	    & ((7, 7), 1397)	& 6	    & \textbf{0.0}		& ((7, 7), 999)\\ 
                                      % 49447	((6, 7), 395)      & 5	  & \textbf{0.0}		& ((0, 1), 315)    	
                                      % 49949	((6, 7), 466)      & 4	  & \textbf{0.0}		& ((7, 6), 347)    	
                                      % 52961	((7, 5), 454)      & 16	  & 6.083   & ((7, 6), 337)    	
                                      % 55973	((5, 7), 242)      & 2	  & \textbf{0.0}		& ((7, 3), 267)    	
                                      % 56977	((7, 6), 457)      & 16	  & 4.083	& ((0, 2), 271)    	
                                      % 57479	((5, 7), 277)      & 16	  & 6.0		& ((7, 7), 278)    	
                                      % 58483	((7, 7), 338)      & 16	  & 4.167	& ((1, 4), 203)    	
                                      % 59989	((7, 7), 252)      & 0	  & \textbf{0.0}		& ((7, 7), 252)    	
                                      % 60491	((7, 7), 237)      & 16	  & 4.0		& ((7, 1), 224)    	
                                      % 63001	((7, 7), 492)      & 4	  & \textbf{0.0}		& ((0, 2), 292)    	
      \midrule
      \multirow{10}{*}{9$\times$8}
      & 100273=509$\times$197    & 8.167	  & ((7, 4), 629)	    & 34	& 4.083	    & ((6, 7), 281)        & 8.083		& ((7, 4), 1413)	& 34	& 8.0		& ((1, 7), 490)     & 4.083		& ((7, 4), 1834)	& 34	& 4.083		& ((1, 7), 754)   \\
      & 101291=509$\times$199    & 8.0	  & ((7, 3), 461)	    & 34	& 6.25	    & ((6, 8), 288)        & 6.083		& ((7, 3), 859)	    & 34	& 8.0		& ((6, 8), 540)     & 8.083		& ((6, 8), 1273)	& 34	& 6.083		& ((5, 8), 692)   \\
      & 107399=509$\times$211    & 8.0	  & ((7, 3), 479)	    & 34	& 4.0		& ((7, 5), 210)        & 4.083		& ((0, 1), 1100)	& 34	& 6.083		& ((7, 6), 431)     & 6.0	    & ((7, 3), 1485)	& 34	& 4.0		& ((1, 4), 701)   \\
      & 113507=509$\times$223    & 8.0	  & ((0, 1), 373)	    & 13	& \textbf{0.0}		& ((7, 6), 244)        & 4.083		& ((0, 1), 1133)	& 3	    & \textbf{0.0}		& ((7, 7), 995)     & 4.083		& ((0, 1), 1803)	& 34	& 6.0		& ((6, 7), 612)   \\
      & 115543=509$\times$227    & 8.083	  & ((0, 1), 541)	    & 34	& 8.0		& ((7, 3), 214)        & 8.0	    & ((0, 1), 1394)	& 34	& 6.167		& ((7, 6), 460)     & 6.0	    & ((0, 1), 1633)	& 34	& 6.083		& ((0, 0), 794)   \\
      & 116561=509$\times$229    & 6.167	  & ((0, 1), 434)	    & 34	& 8.167	    & ((2, 5), 226)        & 6.0	    & ((0, 1), 1002)	& 34	& 6.083		& ((6, 7), 522)     & 6.083		& ((7, 8), 1305)	& 34	& 6.083		& ((7, 8), 660)   \\
      & 118597=509$\times$233    & 6.0	  & ((0, 1), 379)	    & 34	& 4.0		& ((2, 6), 211)        & 8.0	    & ((7, 8), 880)	    & 34	& 6.167		& ((6, 7), 363)     & 6.083		& ((7, 8), 1743)	& 34	& 6.083		& ((5, 8), 683)   \\
      & 121651=509$\times$239    & 8.083	  & ((7, 8), 628)	    & 34	& 4.083	    & ((7, 7), 274)        & 8.0	    & ((7, 8), 1035)	& 9	    & \textbf{0.0}		& ((0, 2), 508)     & 4.0	    & ((0, 1), 1307)	& 4	    & \textbf{0.0}		& ((0, 2), 974)   \\
      & 122669=509$\times$241    & 6.083	  & ((7, 8), 600)	    & 34	& 4.083	    & ((7, 6), 272)        & 10.0		& ((7, 8), 1515)	& 34	& 4.0		& ((0, 4), 557)     & 6.083		& ((7, 8), 2154)	& 34	& 4.0		& ((7, 5), 685)   \\
      & 127759=509$\times$251    & 6.0	  & ((0, 1), 651)	    & 2	    & \textbf{0.0}		& ((0, 1), 542)        & 6.0	    & ((7, 8), 1261)	& 2	    & \textbf{0.0}		& ((7, 8), 1026)    & 4.0	    & ((0, 1), 1799)	& 2	    & \textbf{0.0}		& ((0, 1), 1837)   \\
                                      % 100273	((7, 4), 629)   & 17	& 8.083		& ((7, 5), 339) 	
                                      % 101291	((7, 3), 461)   & 17	& 8.083		& ((6, 8), 287) 	
                                      % 107399	((7, 3), 479)   & 17	& 4.083		& ((0, 2), 207) 	
                                      % 113507	((0, 1), 373)   & 13	& \textbf{0.0}		& ((7, 6), 244) 	
                                      % 115543	((0, 1), 541)   & 17	& 6.167		& ((6, 6), 232) 	
                                      % 116561	((0, 1), 434)   & 17	& 6.167		& ((4, 8), 224) 	
                                      % 118597	((0, 1), 379)   & 17	& 6.167		& ((0, 3), 214) 	
                                      % 121651	((7, 8), 628)   & 17	& 4.167		& ((1, 6), 212) 	
                                      % 122669	((7, 8), 600)   & 17	& 6.167		& ((6, 8), 274) 	
                                      % 127759	((0, 1), 651)   & 2	    & \textbf{0.0}		& ((0, 1), 542) 	
      \midrule
      \multirow{10}{*}{10$\times$8}
      & 201137=1021$\times$197	               & 8.083	 & ((7, 2), 424)	   & 36	    & 10.083	& ((7, 2), 234)         & 2.0	    & ((7, 2), 881)	    & 36	& 4.083		& ((4, 9), 463)      & 4.0	     & ((7, 2), 1430)	& 36	& 6.083	    & ((2, 9), 690)    \\
      & 203179=1021$\times$199	               & 10.083	 & ((0, 1), 474)	   & 36	    & 6.083		& ((0, 2), 281)         & 8.083	    & ((0, 1), 940)	    & 36	& 6.167		& ((5, 7), 423)      & 8.083	 & ((0, 1), 1431)	& 36	& 8.0		& ((7, 6), 1033)    \\
      & 215431=1021$\times$211	               & 8.0	 & ((0, 1), 574)	   & 36	    & 8.0		& ((7, 8), 265)         & 6.0	    & ((0, 1), 1033)	& 36	& 6.083		& ((3, 1), 422)      & 6.0	     & ((0, 1), 1817)	& 36	& 8.0		& ((7, 3), 594)     \\
      & 227683=1021$\times$223	               & 8.083	 & ((0, 1), 586)	   & 36	    & 6.167		& ((5, 9), 213)         & 4.0	    & ((0, 1), 1318)	& 36	& 4.167		& ((6, 8), 419)      & 6.0	     & ((0, 1), 1897)	& 36	& 4.083	    & ((7, 4), 639)   \\
      & 231767=1021$\times$227	               & 10.0	 & ((0, 1), 592)	   & 36	    & 8.083		& ((5, 9), 269)         & 8.083	    & ((0, 1), 1146)	& 36	& 6.083		& ((7, 7), 452)      & 8.083	 & ((0, 1), 1709)	& 36	& 6.0		& ((7, 6), 641)   \\
      & 233809=1021$\times$229	               & 8.083	 & ((7, 9), 361)	   & 36	    & 6.167		& ((5, 9), 248)         & 6.0	    & ((0, 1), 922)	    & 36	& 8.0		& ((2, 9), 453)      & 6.167	 & ((7, 9), 1207)	& 36	& 6.0		& ((7, 5), 776)  \\
      & 237893=1021$\times$233	               & 6.0	 & ((7, 9), 456)	   & 36	    & 6.083		& ((0, 1), 185)         & 6.0	    & ((7, 9), 886)	    & 36	& 6.0		& ((1, 4), 378)      & 4.0	     & ((7, 9), 1480)	& 36	& 6.0		& ((2, 2), 553)  \\
      & 244019=1021$\times$239	               & 8.083	 & ((0, 1), 600)	   & 36	    & 4.0		& ((7, 9), 234)         & 6.167	    & ((0, 1), 1252)	& 36	& 6.0		& ((1, 2), 427)      & 6.083	 & ((7, 9), 1595)	& 30	& \textbf{0.0}		& ((0, 1), 733)   \\
      & 246061=1021$\times$241	               & 2.083	 & ((7, 9), 619)	   & 36	    & 6.0		& ((1, 1), 232)         & 10.0	    & ((7, 9), 1056)	& 36	& 8.0		& ((1, 5), 499)      & 8.0	     & ((7, 9), 1478)	& 36	& 4.0		& ((2, 9), 615)   \\
      & 256271=1021$\times$251	               & 4.083	 & ((7, 9), 659)	   & 36	    & 4.083		& ((7, 8), 226)         & 6.083	    & ((0, 1), 1256)	& 10	& \textbf{0.0}		& ((0, 4), 695)      & 2.083	 & ((0, 1), 1900)	& 36	& 4.083	    & ((7, 8), 787)     \\
      \midrule
  \end{tabular}
\end{sidewaystable}
\FloatBarrier
\newpage
\printbibliography % Print the bibliography

% Can use something like this to put references on a page
% by themselves when using endfloat and the captionsoff option.
\ifCLASSOPTIONcaptionsoff
  \newpage
\fi

% trigger a \newpage just before the given reference
% number - used to balance the columns on the last page
% adjust value as needed - may need to be readjusted if
% the document is modified later
%\IEEEtriggeratref{8}
% The "triggered" command can be changed if desired:
%\IEEEtriggercmd{\enlargethispage{-5in}}

% references section

% can use a bibliography generated by BibTeX as a .bbl file
% BibTeX documentation can be easily obtained at:
% http://mirror.ctan.org/biblio/bibtex/contrib/doc/
% The IEEEtran BibTeX style support page is at:
% http://www.michaelshell.org/tex/ieeetran/bibtex/
%\bibliographystyle{IEEEtran}
% argument is your BibTeX string definitions and bibliography database(s)
%\bibliography{IEEEabrv,../bib/paper}
%
% <OR> manually copy in the resultant .bbl file
% set second argument of \begin to the number of references
% (used to reserve space for the reference number labels box)

% biography section
% 
% If you have an EPS/PDF photo (graphicx package needed) extra braces are
% needed around the contents of the optional argument to biography to prevent
% the LaTeX parser from getting confused when it sees the complicated
% \includegraphics command within an optional argument. (You could create
% your own custom macro containing the \includegraphics command to make things
% simpler here.)
%\begin{IEEEbiography}[{\includegraphics[width=1in,height=1.25in,clip,keepaspectratio]{mshell}}]{Michael Shell}
% or if you just want to reserve a space for a photo:

\end{document}